	\documentclass[preprint,12pt]{elsarticle}
	
	\usepackage{hyperref}

	\usepackage{xcolor}

	\usepackage{siunitx}
        \usepackage{amsmath}
        \usepackage{chngcntr}

	\DeclareMathOperator{\erfc}{erfc}

\counterwithout{equation}{section} 
\counterwithout{equation}{subsection}

\begin{document}
	\begin{frontmatter}

	\title{Test Beam Characterization of a Digital Silicon Photomultiplier}

	\author[]{Finn King\corref{corresAuthor}}
		\cortext[corresAuthor]{Corresponding author}
		\ead{finn.king@desy.de}
	\author[]{Inge Diehl}
	\author[]{Ono Feyens\fnref{vub}}
	\author[]{Ingrid-Maria Gregor\fnref{bonn}}
	\author[]{Karsten Hansen}
	\author[]{Stephan Lachnit}
	\author[]{Frauke Poblotzki}
	\author[]{Daniil Rastorguev\fnref{wupp}}
	\author[]{Simon Spannagel}
	\author[]{Tomas Vanat}
	\author[]{Gianpiero Vignola\fnref{bonn}}
	\address{Deutsches Elektronen-Synchrotron DESY, Notkestr. 85, 22607 Hamburg, Germany}
	\fntext[vub]{Also at Vrije Universiteit Brussel, Belgium}
	\fntext[bonn]{Also at Rheinische Friedrich-Wilhelms-Universit\"at Bonn, Germany}
	\fntext[wupp]{Also at  Bergische Universit\"at Wuppertal, Germany}

	\journal{Nucl. Instrum. Methods Phys. Res. A}

		\begin{abstract}
Conventional silicon photomultipliers (SiPMs) are well established as light detectors with single-photon-detection capability and used throughout high energy physics, medical, and commercial applications. The possibility to produce single photon avalanche diodes (SPADs) in commercial CMOS processes creates the opportunity to combine a matrix of SPADs and an application-specific integrated circuit in the same die. The potential of such digital SiPMs (dSiPMs) is still being explored, while it already is an established technology in certain applications, like light detection and ranging (LiDAR). 

A prototype dSiPM, produced in the LFoundry 150-nm CMOS technology, was designed and tested at DESY. The dSiPM central part is a matrix of 32 by 32 pixels. Each pixel contains four SPADs, a digital front-end, and has an area of $69.6 \times \SI{76}{\square\micro\meter}$. The chip has four time-to-digital converters and includes further circuitry for data serialization and data links.

This work focuses on the characterization of the prototype in an electron beam at the DESY II Test Beam facility, to study its capability as a tracking and timing detector for minimum ionizing particles (MIPs). The MIP detection efficiency is found to be dominated by the fill factor and on the order of \SI{31}{\percent}. The position of the impinging MIPs can be measured with a precision of about \SI{20}{\micro\meter}, and the time of the interaction can be measured with a precision better than \SI{50}{\pico\second} for about \SI{85}{\percent} of the detected events. In addition, laboratory studies on the breakdown voltage, dark count rate, and crosstalk probability, as well as the experimental methods required for the characterization of such a sensor type in a particle beam are presented.
	\end{abstract}


	\end{frontmatter}

		\section{Introduction}\label{sec:introduction}
Silicon photomultipliers (SiPMs) contain an array of single-photon avalanche diodes (SPADs). SPADs are operated in Geiger-mode and sensitive to ionizing radiation and even single photons. One key feature of the Geiger-mode operation are fast-rising signal edges, facilitating single photon time resolutions on the order of~\SI{10}{\pico\second}~\cite{nemallapudi2016}. Structure, operation principle, and applications of SiPMs are discussed, for example, in~\cite{acerbi2019,simon2019,bisogni2019}.

An effect of the Geiger-mode operation is that the SPAD signal contains no information about the energy deposition by the interacting quanta. This inherent digital nature of the SPAD signal makes thinking of a SiPM as a digital device seem natural. As SPAD layouts in CMOS processes are available, CMOS circuitry can be embedded in the SiPM design, which allows profiting from digital signal processing. Introduced as digital SiPMs (dSiPMs) in~\cite{frach2009}, these devices come with several advantages compared to analog SiPMs. The circuitry allows storing information about the position of a firing SPAD (order of \SI{10}{\micro\meter} precision), the masking of noisy SPADs, and on-chip pattern recognition. Parallel connection of the whole SPAD array can be avoided, which reduces the input capacitance to the readout circuit, and thus improves signal rise times. Finally, it can ease system integration, as no separate device for signal digitization is required. One clear disadvantage is the quality of the SPADs, yielding generally higher dark count rates (DCRs) in dSiPMs~\cite{torilla2023}, which will certainly improve once more sophisticated SPAD designs become available in CMOS processes. Another is the reduction of the fill factor, as a part of the surface area is needed for the CMOS circuitry, with increasing circuit complexity.

Possible applications have to find a trade-off between these advantages and disadvantages. Especially, the availability of position information is a key point to consider. One example is to use dSiPMs for the readout of scintillating-fiber bundles~\cite{fischer2022}, where the available position information is used to identify individual fibers, potentially reducing complexity and cost of fiber-readout solutions. Another application would be 4D particle tracking~\cite{cartiglia2022}, where the combination of position information on the order of \SI{10}{\micro\meter} and time information on the order of \SI{10}{\pico\second} might be beneficial properties.

With this perspective, this work reports on the minimum ionizing particle (MIP) detection capabilities found for a new dSiPM prototype designed at DESY~\cite{diehl2023}, and described in section~\ref{sec:sensor}. Laboratory measurements, focusing on breakdown voltage, DCR and crosstalk, are discussed in section~\ref{sec:lab}. Setup, analysis method and results of the MIP-detection studies are presented in section~\ref{sec:setup}, \ref{sec:analysis}, and~\ref{sec:results}, respectively.
	\begin{figure}[]
	\centering
	\includegraphics[width=0.5\textwidth]{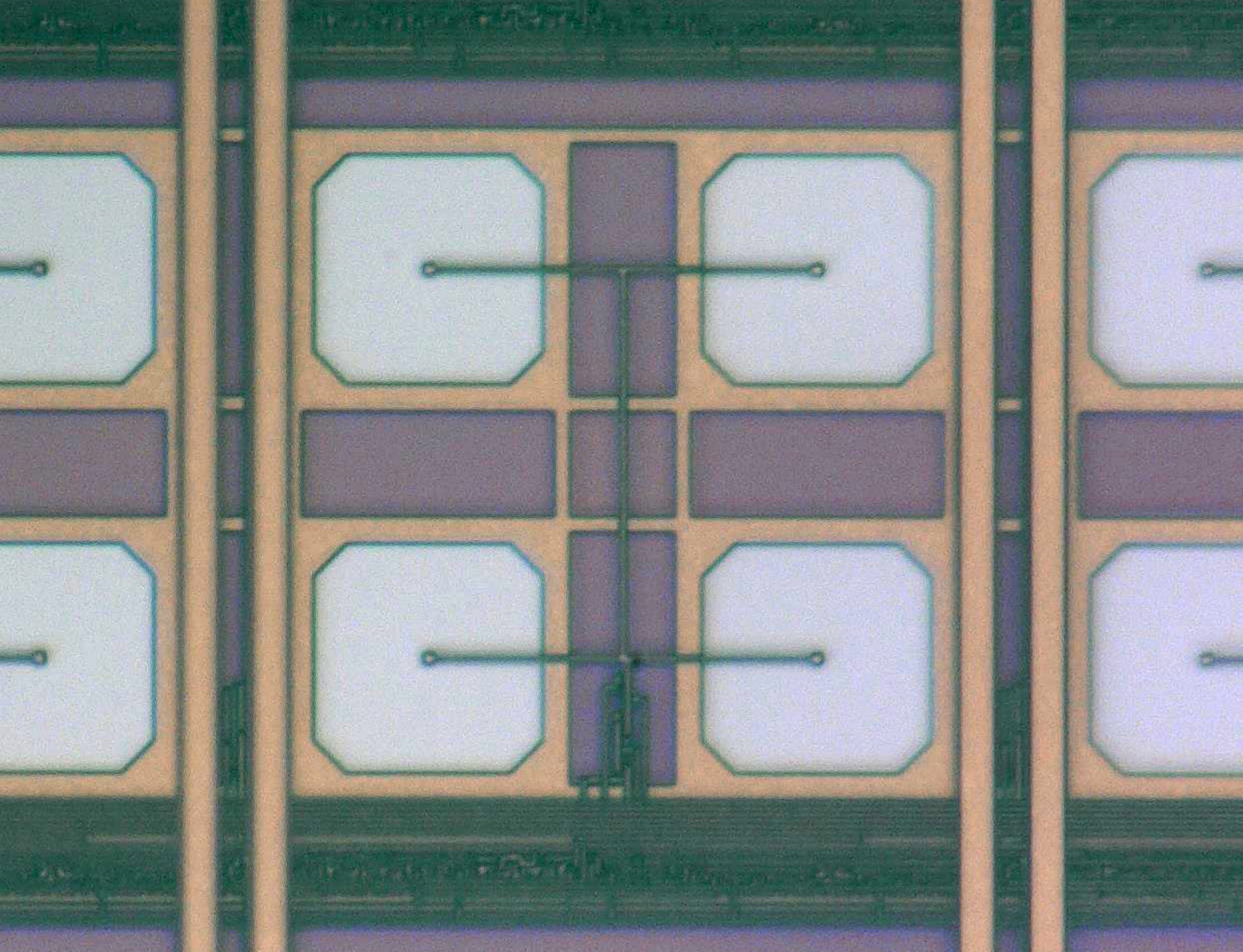}
	\caption[]{Microscopic image of a pixel. The four-SPAD structure with shared electronics in the lower part of the pixel is visible. The total area of the pixel is $69.6 \times \SI{76}{\square\micro\meter}$.}
	\label{fig:pixel}
\end{figure}

	\section{The DESY dSiPM} \label{sec:sensor}
The DESY dSiPM prototype is fabricated in the LFoundry standard 150-\SI{}{\nano\meter} CMOS technology. The basic building blocks of this dSiPM are pixels containing four SPADs with a shared quenching circuit, discriminator and \mbox{2-bit} counter. A microscopic image of the pixel is shown in figure~\ref{fig:pixel}. The pixel area is $69.6 \times \SI{76}{\square\micro\meter}$, where the area occupied by the SPADs amounts to \SI{30}{\percent}, determining the fill factor. A group of $16 \times 16$ pixels forms a quadrant with one 12-bit time-to-digital converter (TDC), generating a time stamp for the first hit (firing SPAD) in each quadrant and readout frame. The readout frames are defined by a \SI{3}{\mega\hertz} clock, and the time and hit position information of each quadrant are buffered and read out in the following frame, if the chip is operated in 1-bit mode. In 2-bit mode, the in-pixel counter is used to store up to three hits per pixel and frame. Consequently, the data transmission takes two frames in this case. For the studies presented hereafter, the 1-bit mode is used. Individual pixels can be disabled (masked) using a transistor switch in series with the quenching transistor, which helps to reduce the DCR in the presence of noisy pixels. A temperature diode in the periphery of the chip allows for precise temperature monitoring during measurements and operation.

Further features of the DESY dSiPM, like a validation logic for topology-based event selection, external test circuits, and LED trigger pulse generation, are detailed in~\cite{diehl2023}, along with a detailed report on the ASIC design and properties.

		\subsection{Data Acquisition System}
To facilitate data acquisition (DAQ) and save development time, the DESY dSiPM is integrated into the Caribou DAQ system, an open-source hardware, software (Peary) and firmware project dedicated to pixel detector development~\cite{vanat2020,caribou}. Its key component is a system-on-chip (SoC) board (AMD~\cite{amd}, Zynq 7000 SoC ZC706) containing an FPGA and a CPU. The CPU is running a fully-featured Yocto-based Linux distribution and Peary, while the FPGA executes custom written firmware blocks for configuration, control, and readout of the dSiPM. The SoC board is connected to a mezzanine board, the CaR board, that provides supplies like current and voltage sources, and a physical interface to the dSiPM. The dSiPM itself is placed on a custom chip board, connected to the CaR board, and hosting LVDS repeaters, for communication with the chip, and further interfacing circuits.

The Caribou system provides everything required for operation of the dSiPM, but the high voltage required for biasing of SPADs, which requires an external voltage source. The provided supplies include power, current references, voltage references, LVDS links, a \SI{3}{\mega\hertz} clock, a \SI{408}{\mega\hertz} clock, and synchronous control signals generated in the FPGA. This way, the entire system is relatively compact, which eases physical integration at the test beam, in a laser setup, or for microscopy.

The Peary software provides an interface to control the supplies on the CaR board, and to registers in the FPGA for control of the firmware, and data transactions. It also provides a development environment to define sensor-specific powering, configuration and readout procedures. Following a certain structure, naming convention, and implementing key functional features, these procedures are intrinsically embedded in the EUDAQ data acquisition software framework~\cite{ahlburg2020,EUDAQ}, used with the test-beam instrumentation introduced in section~\ref{sec:setup}.

		\section{Laboratory Characterization} \label{sec:lab}
Details on the laboratory characterization of the DESY dSiPM --- with focus on digital properties --- are reported in~\cite{diehl2023}. This section discusses breakdown voltage, and crosstalk measurements, to give additional insights in properties of the dSiPM and to enable the operation at comparable overvoltages.
\begin{figure}[tbp]
	\centering
	\includegraphics[width=0.48\textwidth]{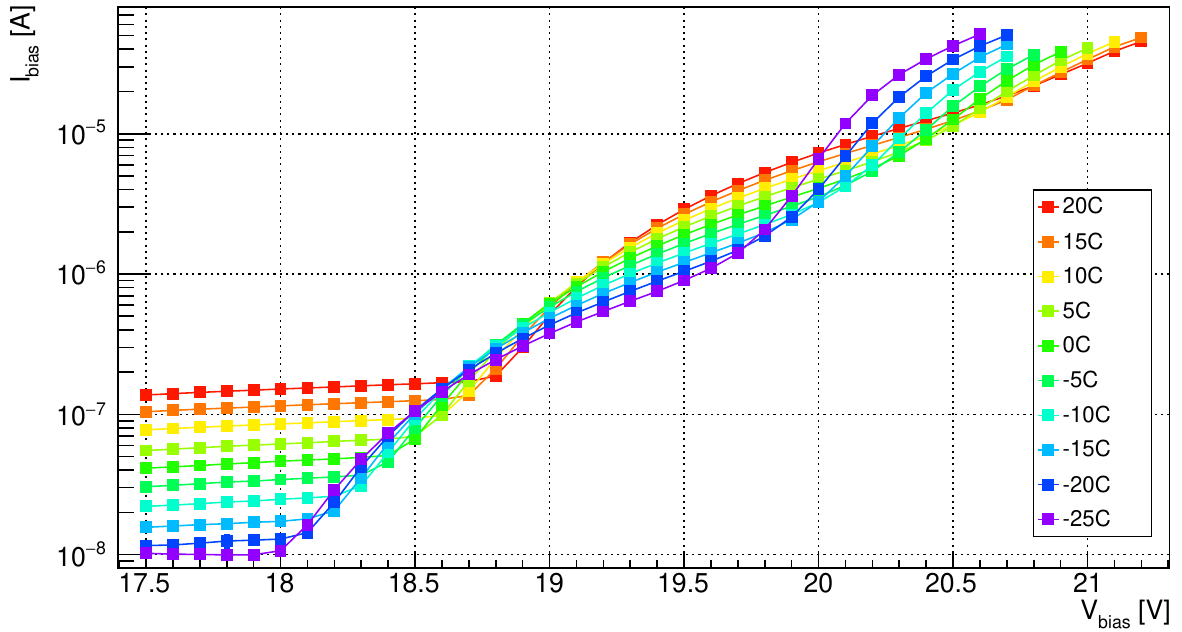}
	\caption[]{IV-curves for one of the tested sensors, at temperatures between~\SI{-25}{\degreeCelsius} and~\SI{20}{\degreeCelsius}. Each point represents the average of several measurements; the errors are within the marker.}
	\label{fig:IV}
\end{figure}

\subsection{Breakdown Voltage Measurements}
The breakdown voltage $V_{BD}$ marks the reverse bias voltage, at which SPADs start to exhibit avalanche multiplication leading to Geiger discharges. This is an important characteristic of a SiPM, since they are operated at an overvoltage $V_{OV}$ above the breakdown voltage. As summarized in~\cite{klanner2019}, the breakdown voltage can be extracted from a measurement of the bias current $I_{bias}$ of a SiPM as a function of the reverse bias voltage $V_{bias}$ (IV-curve).

Measurements of IV-curves are made inside a climate chamber (ACS~\cite{acs}, model DY200), assuring a temperature and humidity controlled environment and dark conditions. A source meter (Keithley~\cite{keithley}, model 2410) is used for fine control of reverse bias and precise current measurement. Figure \ref{fig:IV} shows IV-curves for one of the tested sensors, at environmental temperatures between~\SI{-25}{\degreeCelsius} and~\SI{20}{\degreeCelsius}. As can be seen, breakdown, marked by the sharp rise of the current, occurs between \SI{18}{\volt} and \SI{19}{\volt} and the current before breakdown is between \SI{10}{\nano\ampere} and \SI{200}{\nano\ampere}, both depending on temperature. It should be noted that this is a purely analog measurement and does not depend on the digital characteristics of the sensor but rather on SPAD design and manufacturing process.

\begin{figure}[tbp]
	\centering
	\includegraphics[width=0.48\textwidth]{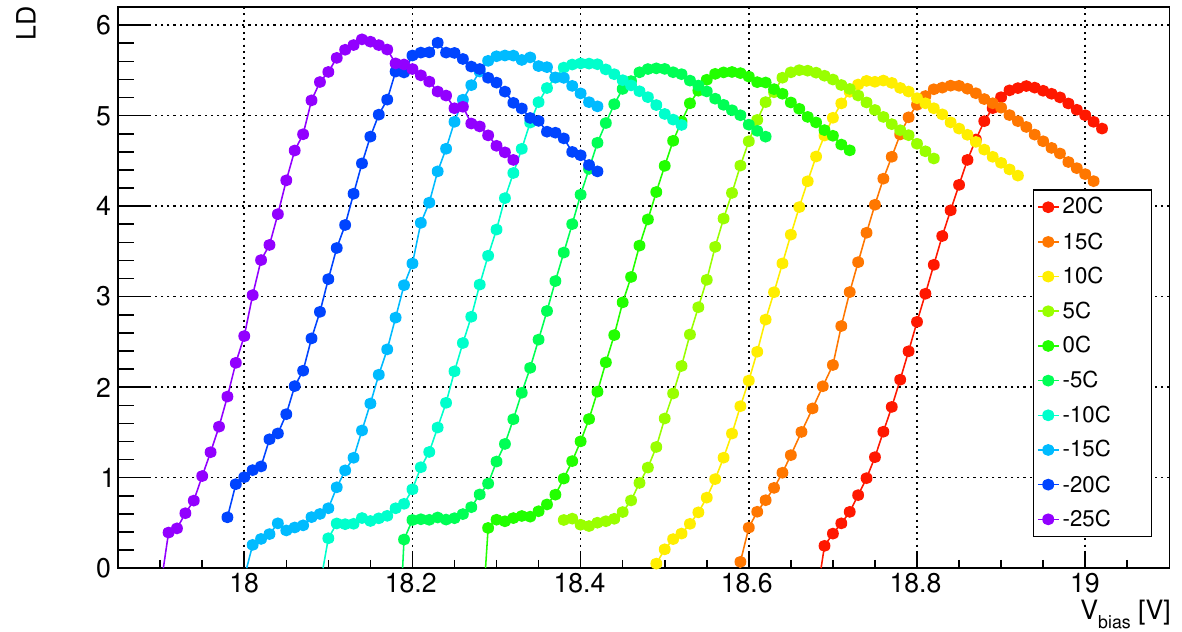}
	\caption[]{Logarithmic derivatives of IV-curves, for one of the tested sensors, at temperatures between~\SI{-25}{\degreeCelsius} and~\SI{20}{\degreeCelsius}.}
	\label{fig:BD}
\end{figure}

Several methods to quantify $V_{BD}$ are based on identifying the previously mentioned sharp rise in the IV-curve. Some of these methods are compared in~\cite{klanner2019}. Within this work, the breakdown voltage is defined as the maximum of the logarithmic derivative
\begin{equation}
\label{eq:RD}
	LD = \frac{d}{dV_{bias}} ln(I_{bias}),
\end{equation}
which proved to provide a robust estimate, and is reported to be in good agreement with other methods. A set of IV-curves for the same temperatures, and with finer voltage steps, is taken around the breakdown voltage estimated from figure~\ref{fig:IV}. The corresponding logarithmic derivative is shown in figure~\ref{fig:BD}, and the estimated breakdown voltages are e.g. \SI{18.14}{\volt} and \SI{18.93}{\volt}, at~\SI{-25}{\degreeCelsius}, and~\SI{20}{\degreeCelsius}, respectively. These measurements are performed for all investigated samples, so that consistent overvoltages for the following studies can be chosen in the relevant temperature range. It should be noted that the logarithmic derivative is calculated between points with a distance $dV_{bias} = \SI{0.08}{\volt}$. This choice has a systematic impact on the measured breakdown voltage, but yields a robust estimate, and is the same for all measurements.

\subsection{Crosstalk Measurements}
The Geiger discharge within a SPAD generates secondary photons, which can generate secondary electron-hole pairs in adjacent SPADs and thus trigger additional discharges. This effect is known as optical crosstalk, and it occurs with a probability proportional to the gain of the SPAD~\cite{xtalk2013}. As discussed in~\cite{klanner2019,acerbi2019}, one can differentiate prompt and delayed optical crosstalk, where the secondary charge carriers are generated in the depleted, and non-depleted region, respectively. The crosstalk probability can be reduced by means of trenches filled with optically dense material between SPADs~\cite{trench2009}.

While this effect is considered a nuisance to photon counting applications, it could be beneficial for MIP detection. This is because the affected SPAD cell might be in another pixel and hence indicate a particle interaction close to the boundary of the pixels, similar to charge sharing in silicon pixel or strip sensors. Crosstalk measurements on a single pixel level are discussed in the following, while the macroscopic effects of crosstalk are discussed in~\ref{sec:results:res}.

\subsubsection{Single Pixel Crosstalk}
As the DESY dSiPM provides information about the position of a firing pixel, the crosstalk probability can be measured for individual pixels using an approach similar to the one described in~\cite{liu2016}. First, an isolated noisy pixel is chosen. In step 1, the isolated DCR of this pixel, and its eight neighbors, is measured successively by masking the entire matrix but the studied pixel. One should note, that this eliminates crosstalk contributions from neighboring pixels, as the masking mechanism prevents Geiger discharges in the masked pixels. In step two, the central pixel plus one of the neighbors are unmasked and the DCR measurement is repeated, again successively for all eight neighbors. The total recorded number of frames $N$ and the number of frames $N_c$, $N_n$, where the central, or a neighboring pixel fires in step one, allow estimating the probabilities
\begin{align}
	p_c &= \dfrac{N_c}{N}, \text{and} \\
	p_n &= \dfrac{N_n}{N}.
\end{align}
From the number of frames where both pixels fire in step two, $N_{c \land n}$, one can derive
\begin{align}
	p_{c \land n} &= \dfrac{N_{c \land n}}{N}.
\end{align}
This allows to derive the crosstalk probability $p_x$ to a given neighbor
\begin{align}
p_x = \dfrac{p_{c \land n} - p_c p_n}{p_c(1-p_n)}, 
\end{align}
assuming that $p_c$ is significantly larger than $p_n$.

\begin{figure}[tbp]
	\centering
	\includegraphics[width=0.48\textwidth]{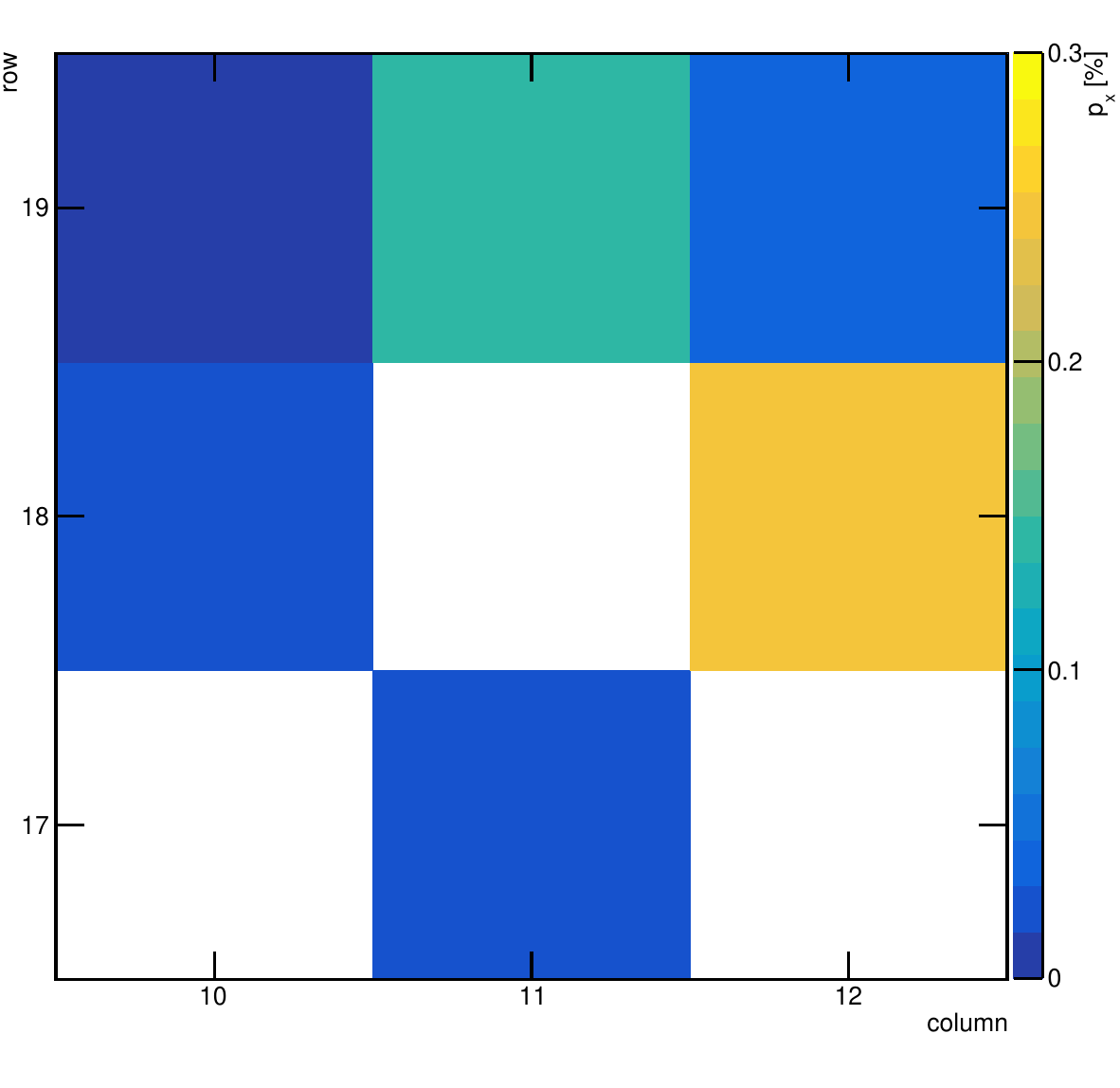}
	\caption[]{Measurement of the probability of crosstalk between a noisy pixel (center), and its neighbors. Measured at an overvoltage of about \SI{2}{\volt} and a temperature of about \SI{0}{\degreeCelsius}. Count rates in the outermost pixels of the bottom row are too low for an estimate of the crosstalk probability.}
	\label{fig:CT}
\end{figure}

An example for one pixel, at an overvoltage of about \SI{2}{\volt}, and a sensor temperature of about \SI{0}{\degreeCelsius}, is shown in figure~\ref{fig:CT}. It can be seen that the probability of crosstalk is non-uniform, and largest for the pixel right of the central one. It is observed, that the total probability of crosstalk and the position of the pixel with the largest crosstalk, vary between several studied noisy pixels. The arrangement of the SPADs inside the pixel, see figure~\ref{fig:pixel}, suggests that a certain SPAD within the pixel is responsible for the increased DCR, and consequently crosstalk to proximate pixels is more likely. Assuming that a localized production defect determines the increased DCR, even the position of this defect within the noisy SPAD will impact the measured crosstalk probability.

\begin{figure}[tbp]
	\centering
	\includegraphics[width=0.48\textwidth]{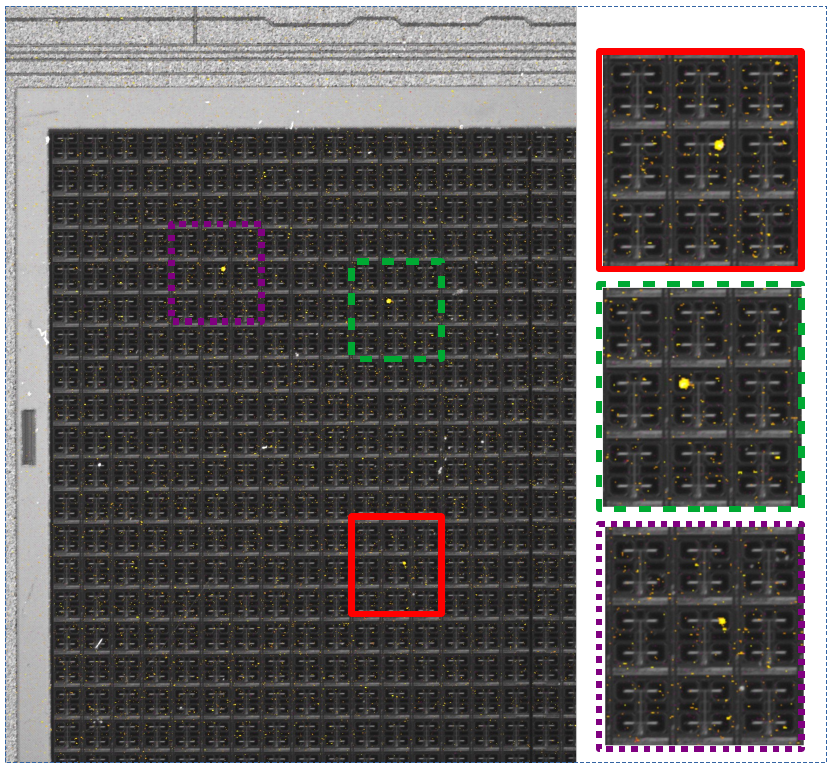}
	\caption[]{Measurement of the light emitted in the avalanches process in noisy pixels. The image, in shades of yellow, is superimposed to a microscopic image of the matrix, displayed in grayscale. Close-ups of the region around the three noisiest pixels are shown on the right side of the image. The solid red box indicates the group of pixels shown in figure~\ref{fig:CT}.}
	\label{fig:light}
\end{figure}

To confirm this, the emission of visible light, generated during the Geiger discharge in noisy pixels, is measured in dark conditions. This method is based on~\cite{dcr_lumi}, and performed on the same pixel, using a microscope (Zeiss~\cite{zeiss}, Axioscope), and a CCD camera (Jenoptik\cite{jenoptik}, model ProgRes SpeedXT core5). Figure~\ref{fig:light} shows microscopic pictures of the pixel matrix, super-imposed with the obtained image. The image is recorded with an exposure time of \SI{50}{\second}, at the maximum overvoltage, the lowest resistance of the quenching transistor, and corrected for dark noise from the CCD sensor to optimize the signal-to-noise ratio. A clear spot of emitted light is visible close to the top-right SPAD of the pixel marked in figure~\ref{fig:CT}, confirming the hypothesis from above. One should note, that also some difference in crosstalk to the top and right neighbor is expected, as the placement of the in-pixel electronics on the bottom of a pixel results in a larger distance to the closest SPAD in the neighboring pixel. Further noisy pixels can be identified in the image, and the measurement of crosstalk to their neighbors yields consistent results.

The studies show that a crosstalk probability on percent level is to be expected for typical operation conditions, which is too small to have a significant effect on the spatial resolution. It also shows, that the DCR of noisy pixels is generally dominated by single noisy SPADs. Thus, a masking mechanism with higher granularity (than an entire pixel) would be beneficial, as a similar reduction in DCR would require only \SI{25}{\percent} of the masked area. One should note that the shown crosstalk probability is not corrected for the occurrence of more than one dark count per frame (relevant at high DCR, or due to after-pulsing), and that delayed crosstalk events are not necessarily recorded in the same frame.

		\section{Test-Beam Setup} \label{sec:setup}
The presented test-beam measurements follow a simple concept. A collimated beam of electrons penetrates the device under test (DUT), in this case two DESY dSiPMs. A set of reference detectors provides trigger signals and tracking on a single electron level. This way, electron trajectories (tracks) can be reconstructed during (online) or after (offline) beam operation (run). This allows to reconstruct the position at which an electron hits the DUT and to characterize the DUT in terms of hit detection efficiency, and temporal as well as spatial hit resolution. The hit efficiency denotes the probability to detect MIPs. The spatial resolution refers to the uncertainty of the position measurement provided by the DUT, the temporal resolution is defined in the same way. The reconstruction of these properties is discussed in section~\ref{sec:analysis}.

\begin{figure}[tbp]
	\centering
	\includegraphics[width=0.5\textwidth]{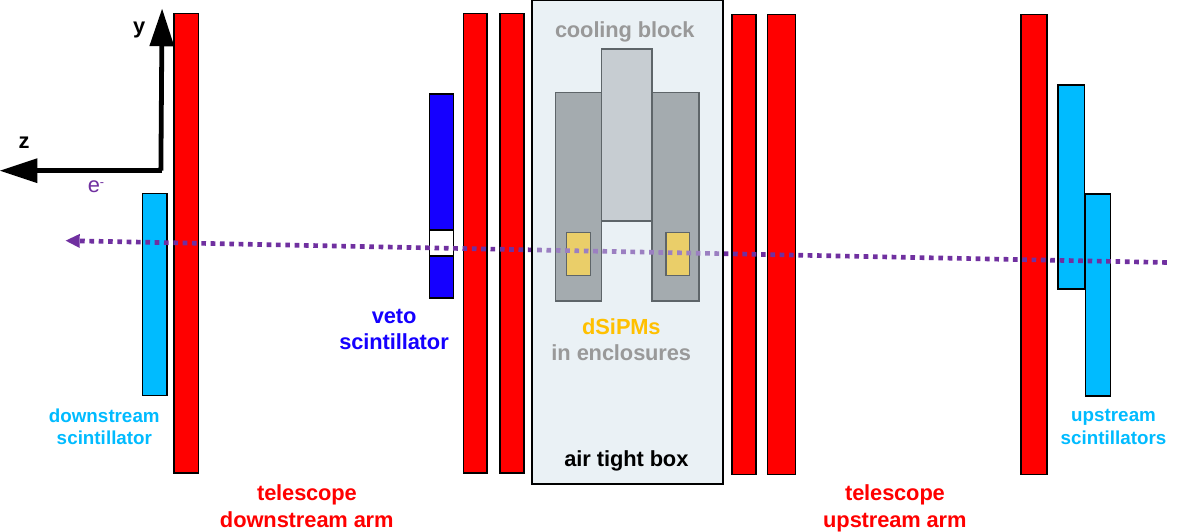}
	\caption[]{Schematic image of the test-beam setup, showing the position of the DESY dSiPMs with respect to the reference detectors for trigger signals (scintillators) and tracking (telescope).}
	\label{fig:setup}
\end{figure}

The test-beam measurements were conducted in two weeks, in March 2023, at the  DESY II Test Team Facility~\cite{desyii}. An electron momentum of \SI{4}{\giga\electronvolt\per c} is selected as a trade-off between electron rate and track resolution, which is affected by multiple Coulomb scattering~\cite{moliere,bethe}. A schematic image of the test-beam setup is shown in figure~\ref{fig:setup}. The set of reference detectors for particle tracking is the so-called DATURA beam telescope~\cite{eudet}. It consists of six MIMOSA 26 sensors~\cite{mimosa} with an intrinsic spatial resolution of \SI{3.24}{\micro\meter}. The sensors are arranged in two groups of three, with a spacing between about \SI{20}{\milli\meter} and \SI{300}{\milli\meter}. The trigger system consists of four scintillators coupled to photomultiplier tubes. A coincidence logic between three of the scintillators is used to suppress noise. A fourth scintillator features a hole of about \SI{2}{\milli\meter} diameter, which is slightly smaller than the active area of the dSiPMs. It is placed downstream of the dSiPMs, as it adds no material in the area of interest, and used as a veto, to discard tracks outside the active area of the dSiPMs. The other three scintillators are placed outside the beam telescope, so that the additional material does not deteriorate the tracking resolution.

To facilitate synchronous operation of the dSiPMs, and the trigger and tracking system, an AIDA trigger logic unit (TLU) is used~\cite{tlu}. The TLU derives the trigger signal from the scintillator signals and distributes it to the other detectors in the setup. To ensure that the DAQ systems of the beam telescope and the dSiPMs are ready to accept a trigger, and not occupied processing a previous trigger signal, both devices provide a busy signal to the TLU, which vetoes the generation of trigger output signals. In addition, the TLU provides a clock and a run start signal to ensure synchronous operation of the detector systems. The firmware running on Caribou continuously buffers the data packets for each readout frame of the dSiPMs. As the trigger signals arrive with a certain delay, and asynchronously with respect to the \SI{3}{\mega\hertz} frame clock, the corresponding hit might be registered in the current or previous readout frame. Thus, both frames are marked for storage, and combined for the later analysis. In some measurements, more than two frames are marked for storage, to have data for noise studies under the conditions at the test beam.

\begin{figure}[tbp]
	\centering
	\includegraphics[width=0.5\textwidth]{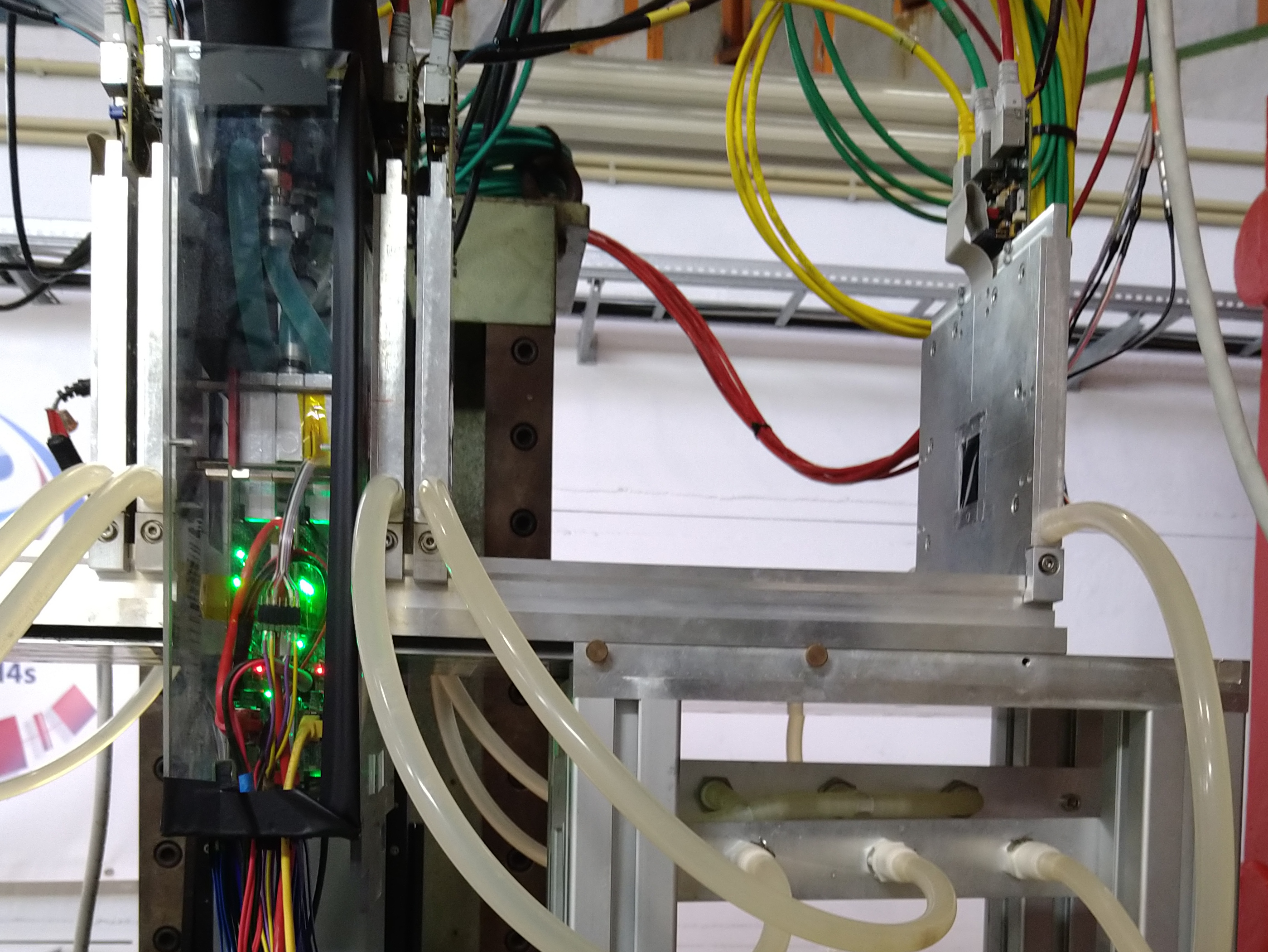}
	\caption[]{Test-beam setup used in March 2023. The transparent plastic on the left contains the dSiPMs in their aluminum casing, the CaR board, and tubes for the light blue coolant. The five aluminum casings on the left and right contain the MIMOSA 26 sensors.}
	\label{fig:setup_photo}
\end{figure}

The chip boards with the dSiPMs are placed inside aluminum casings, which provide light shielding and thermal contact to a cooling element. This allows stabilizing the operational temperature at about \SI{0}{\degreeCelsius}, limited by the power of the cooling unit and losses. The aluminum casings and cooling elements are inside a nitrogen flushed plastic casing, as shown in figure~\ref{fig:setup_photo}. The backplate of the casing is mounted on a set of position stages to allow movement orthogonal to the beam direction with micrometer precision. The chip boards, the aluminum casings, and the surrounding plastic casing feature holes to minimize the amount of material affecting tracking precision. Thin layers of aluminum or Kapton foil were used to keep the casings light- and airtight.

One should note that the trigger area, given by the hole in one of the scintillators, is slightly smaller than the active area of the dSiPM. This is a compromise to maximize the fraction of triggers that actually hit the DUT. On the other hand, it requires the mechanical alignment of the DUT and the trigger system to be on the order of \SI{0.1}{\milli\meter}. The mechanical alignment of the trigger system relative to the beam telescope is straight forward, as the trigger area is easily recognizable in the hit maps of the MIMOSA sensors. If the DUT were able to providing a trigger output itself, it could be aligned in the same way. Unfortunately, the high DCR does not permit this approach. Instead, an approach based on the evaluation of the electron scattering in the material around the DUT is applied, as presented in section~\ref{sec:analysis} and~\ref{sec:results}.

		\section{Analysis Procedure} \label{sec:analysis}
The analysis of the test-beam data is performed with the Corryvreckan framework~\cite{corry,corry_web}. This software analyzes the raw data from the TLU, MIMOSA sensors, and DESY dSiPMs, allowing for the reconstruction of particle trajectories and observables quantifying the dSiPMs MIP-detection capabilities, within an event-based structure.

As a first reconstruction step, the raw data are decoded and the hit information is associated to events, which are defined by TLU trigger IDs. Next, clusters are reconstructed as groups of neighboring hits in the telescope planes and the dSiPMs (merging data from both readout frames associated with the given trigger ID). The cluster center is calculated as the mean position of all pixels in the cluster and estimates the point at which the electron traversed the corresponding detector. Electron tracks are reconstructed from the cluster centers in the telescope planes, applying the General-Broken-Lines track model~\cite{gbl1,gbl2}, which allows for kink angles in the particle trajectory at scatterer positions (telescope planes, DUT, etc.). To maximize tracking precision, only tracks with an associated cluster on each telescope plane are analyzed, and tracks with $\chi^2 / n_\text{dof} > 5$ are rejected. Also, tracks outside a region of interest related to the trigger acceptance are discarded. Hits on the dSiPMs are associated to the reconstructed tracks when they are within a radius of \SI{70}{\micro\meter} around the reconstructed intersection of the particle track and the corresponding dSiPM plane. Additionally, the time stamps of the dSiPM hits are requested to be within \SI{230}{\micro\second} around the time stamp of the corresponding TLU trigger, to account for the integration time of the MIMOSA 26 sensors.

The presented measurements and their analysis required some uncommon procedures for physical and software alignment, described in section~\ref{sec:analysis:align}. The investigated observables, quantifying the dSiPMs MIP-detection performance, are introduced in section~\ref{sec:analysis:obs}.

		\subsection{Physical and Software Alignment} \label{sec:analysis:align}
The analysis of test-beam data relies on precise knowledge of the relative positions and orientations of the involved detectors. This is achieved using dedicated alignment modules within the Corryvreckan framework, which are based on minimizing the global $\chi^2$ of all tracks, a standard procedure in testing particle-tracking detectors.

Given the small size of the hole in the trigger scintillator, defining the acceptance for electron tracks, mechanical alignment of the dSiPMs with respect to the reference detectors needs to be on the order of \SI{0.1}{\milli\meter}. This needs to be done when setting up, before starting final measurements. Typically, this is achieved by using a trigger signal derived from the response of the DUT, and analyzing the position and shape of the corresponding acceptance on a close telescope plane. For the dSiPM, this approach is not applicable, as there are dark counts in the majority of the frames. Hence, the relation between the electron impact position and the presence of a trigger signal is not given.

Instead, measurements at an electron momentum of \SI{2}{\giga\electronvolt\per c} are recorded, using the coincidence of two scintillators without hole as a trigger input. This is to increase scattering effects and the investigated area, respectively, otherwise the setup remains unchanged. For the analysis of the acquired data, Corryvreckan is configured to perform so called material-budget imaging (MBI), using a module developed in the context of~\cite{ect}. This approach assumes that the dominating amount of scattering material lies in the plane of a single DUT, between the upstream and downstream arm of the telescope (see figure~\ref{fig:setup}). Straight lines are reconstructed in the upstream and downstream arm, separately. Pairs of tracks in both arms are defined, applying a matching cut of \SI{200}{\micro\meter} on the intercept of the tracks with the DUT plane. This allows to derive a scattering angle (kink) between the upstream and downstream track.

\begin{figure}[tbp]
	\centering
	\includegraphics[width=0.48\textwidth]{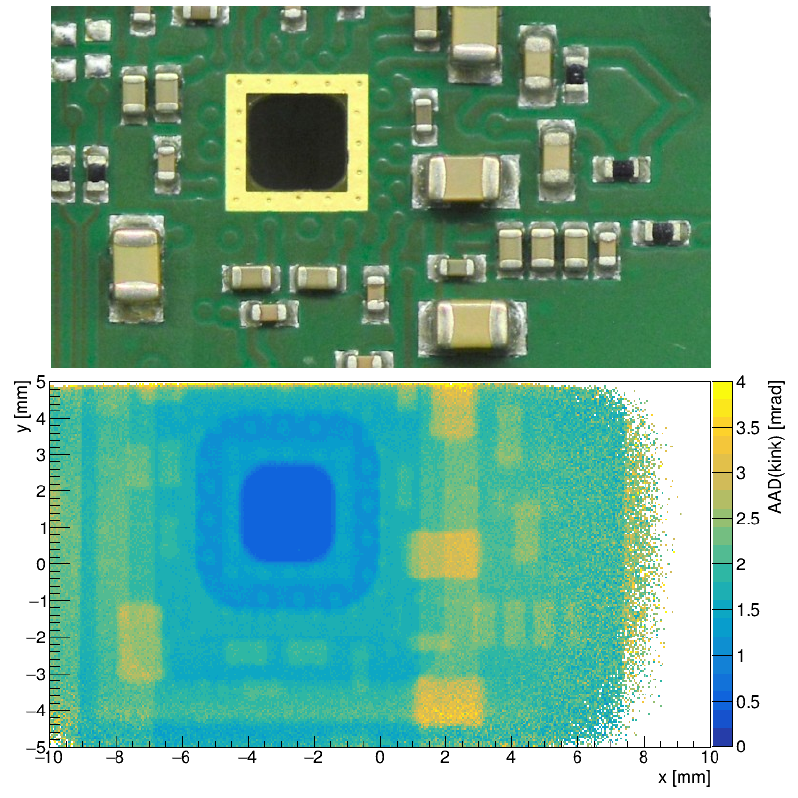}
	\caption[]{The upper part shows a microscopic picture of the dSiPM chip board. The dSiPM is placed on the backside of the board, behind the hole in its center. The lower part shows the width of the scattering-angle distribution as a function of position in the DUT plane.}
	\label{fig:MBI}
\end{figure}

The width of the distribution of scattering angles is related to the traversed material budget~\cite{moliere,highland}. Calculating the average absolute deviation (AAD) of the kink angle as a function of position in the DUT plane yields the result shown in figure~\ref{fig:MBI}. It allows identifying the position of the dSiPM relative to the reference system by identifying the hole in the chip board and characteristic components, also shown in the figure. It also shows, that the two DUTs are well aligned with respect to each other.

\begin{figure}[tbp]
	\centering
	\includegraphics[width=0.48\textwidth]{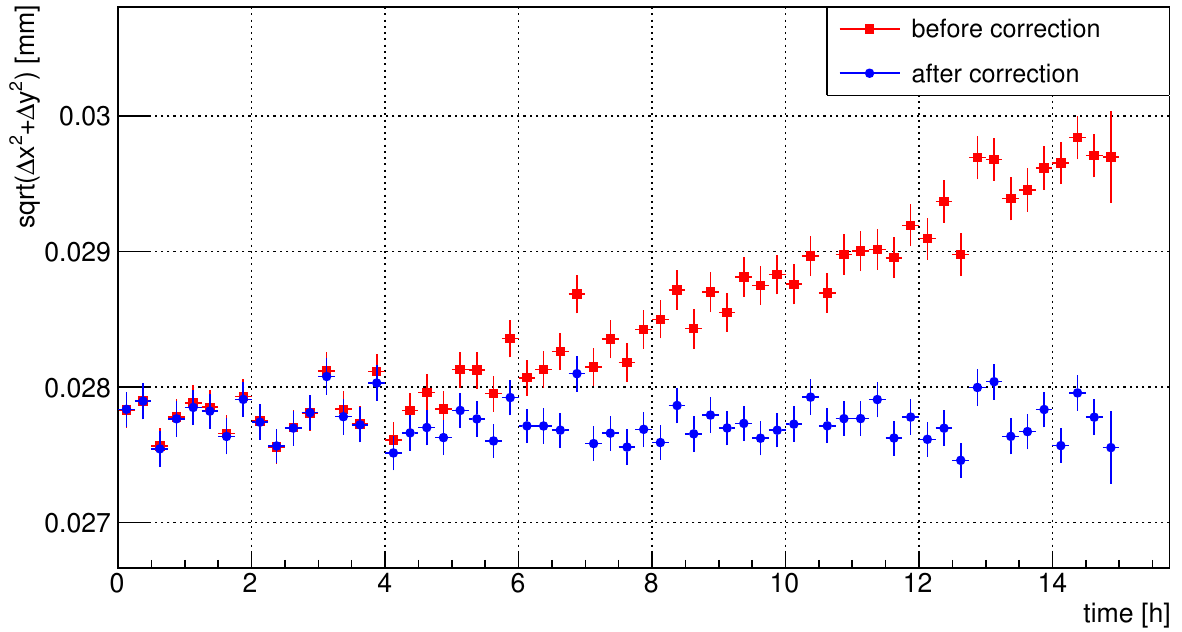}
	\caption[]{The root-mean-square of the residuals in x- and y-direction is before and after applying the correction for thermal expansion described in the text.}
	\label{fig:RSS}
\end{figure}

Offline analysis of the data revealed movements of the dSiPM, relative to the reference system, on the scale of a few \SI{}{\micro\meter\per\hour}. This is assumed to be due to thermal expansion of mechanical components. To correct for this effect, the alignment of the DUT is performed using a part of the data at the beginning of a run, where relative movement can be considered negligible. The entire data of the run is analyzed in a next step, using the previously found alignment parameter. The spatial residuals, defined in section~\ref{sec:analysis:obs:res}, are then analyzed as a function of time. A polynomial fit to the obtained data is then used to parameterize the time dependence of the alignment parameters, and used as correction in the resolution studies. Figure \ref{fig:RSS} shows the effect of the described correction on the root-sum-square of the spatial residuals in x- and y-direction.

\subsection{Observables} \label{sec:analysis:obs}
To investigate the MIP-detection performance of the dSiPMs, clusters on the dSiPMs are associated with the reconstructed tracks, as described in the beginning of this section. The set of selection criteria is the same for all observables, unless mentioned otherwise.

\subsubsection{Efficiency} \label{sec:analysis:obs:eff}
The hit-detection efficiency of a tracking detector denotes the probability of identifying a signal, when a MIP penetrates the pixel matrix of a given sensor. In a test-beam experiment, it is reconstructed as the ratio
\begin{equation} \label{eq:efficiency_rec}
\epsilon_{\text{{rec}}} = \frac{N_{\text{hits}}}{N_{\text{tracks}}},
\end{equation}
where $N_{\text{tracks}}$ denotes the number of tracks passing the selection criteria, and $N_{\text{hits}}$ is the subset of tracks with an associated cluster on the DUT.

The relatively high DRC of the dSiPMs causes some additional corrections to be required, to prevent a bias to the quoted hit-detection efficiency. First, tracks that intercept a masked pixel, or one of its direct neighbors, are discarded. Pixels are masked when their hit rate is about two orders of magnitudes larger than the average. Still, the reconstructed efficiency $\epsilon_{\text{rec}}$ needs to be corrected for the non-negligible probability of associating fake hits with the reconstructed track. To account for the strong dependency of the DCR on operation conditions, the density of fake hits is estimated from the same data, defining clusters as fake when there is no track within a radius of 2 pixel-pitches around them. Normalizing to the area defined by the association cut yields the fake-hit probabilities for the two readout frames $f_{F1}$ and $f_{F2}$. This allows to define a corrected efficiency

\begin{equation} \label{eq:efficiency}
\epsilon = c_{\text{dt}} \cdot \frac{{\epsilon_{\text{{rec}}} - (f_{F1} + f_{F2} - f_{F1} f_{F2})}}{{1 - (f_{F1} + f_{F2} - f_{F1} f_{F2})}},
\end{equation}
where the factor $c_{\text{dt}} = 1.0342$ corrects for the dead time between the two readout frames \footnote{Each readout frame has a duration of \SI{333}{\nano\second}. However, due to the readout scheme described in ~\cite{diehl2023}, hits within a reset phase of \SI{11}{\nano\second} cannot be detected. It still happens, that the reference system issues trigger signals for particles interacting during that reset phase. The given correction factor corresponds to the ratio $333 / (333-11)$.}.

The method to correct for the noise contamination was cross-validated by acquiring an additional 3rd frame, that does not contain hits due to particle interactions, as it is too late in time. Applying the described analysis procedure only on data from this third frame, the per-frame fake-hit contribution to the efficiency $f_{F3}$ can be estimated directly. Replacing $f_{F1}$ and $f_{F2}$ in equation~\ref{eq:efficiency} with $f_{F3}$, allows applying the correction in a similar way. The results from these two methods are found to be compatible within the uncertainties.

\subsubsection{Spatial Residuals and Resolution} \label{sec:analysis:obs:res}

The spatial resolution of a tracking detector is given by the error on its measurement of a MIPs impact position. This can be reconstructed from the residuals
\begin{equation} \label{eq:residual}
dx = x_{\text{track}} - x_{\text{cluster}},
\end{equation}
where $x_{\text{hit}}$ is the reconstructed position of an associated cluster in a dSiPM, and $x_{\text{track}}$ is the position at which the reconstructed particle track intercepts this dSiPM. Note that the coordinates are given in a Cartesian coordinate system, where the z-axis is orthogonal to the sensor plane, and the x- and y-axes are parallel to the column- and row-direction of the dSiPMs. The width of the distribution $dx$ is determined by the uncertainties of $x_{\text{track}}$, and $x_{\text{cluster}}$, respectively denoted as $\sigma_{\text{x,hit}}$, and $\sigma_{\text{x,track}}$. The tracking uncertainty $\sigma_{\text{x,track}}$ depends on the electron momentum, geometrical setup, and amount of material the electron needs to penetrate. It is estimated to be between \SI{3.3}{\micro\meter} and \SI{3.7}{\micro\meter} at the positions of the DUTs, using the General Broken Lines formalism~\cite{res_calc} for the given setup and conditions.

The distribution of $dx$ is determined by the shape of a signal and a background contribution, denoted $S(dx)$ and $B(dx)$, respectively. The sum of these parametrizations is fitted to the reconstructed distributions, to extract the spatial resolution without impact from background. The signal contribution is determined by the positions of the SPADs within the pixel cell (see figure~\ref{fig:pixel}), as they define the efficient part of the pixel. It is parameterized as the normalized sum of four error functions,
\begin{equation}
\label{eq:Signal}
S(x) = N_S \cdot \frac{f_1-f_2+f_3-f_4}{2},
\end{equation}
where
\begin{equation*}
f_{i}(x) = \text{{erf}} \left(\frac{{x \pm C \pm \frac{W}{2}}}{{\sqrt{2}\sigma}}\right).
\end{equation*}
The parameters $C$, and $W$ represent the position of the SPAD center, fixed to be at \SI{17.4}{\micro\meter}, and the nominal active width of the SPAD, fixed at \SI{19.93}{\micro\meter}, respectively. The free parameters, $N_S$ and $\sigma$, represent a normalization factor and the width of the Gaussian smearing, which includes the tracking uncertainty and the gradual decline of the efficiency at the boundaries of the SPADs.

The background contribution, on the other hand, stems from erroneously associating a noise hit (dark count) with a track, that was not registered as a hit in any pixel. The noise contamination has been reduced through online or offline masking of noisy pixels, but it cannot be completely eliminated due to the typical properties of SPADs, especially at high overvoltages and temperatures. The shape of the background contribution is obtained integrating the acceptance of the radial cluster-association cut, and yields
\begin{equation}
\label{eq:Noise}
B(x) = N_B \cdot 2 \cdot \left( \sqrt{{\left| D^2 - x^2 \right|}} \right),
\end{equation}
where $D$ represents the radial cut, and $N_B$ is a normalization factor and a free parameter of the fit.

Fitting the combination of the signal and the background distribution, allows their separation for an unbiased analysis of the dSiPMs spatial hit resolution.

\subsubsection{Cluster Size}
The cluster size is defined as the number of pixels contributing to a cluster, and reconstructed for all clusters selected for the spatial residuals. It exceeds one for certain types of sensors and operation conditions, mainly because of charge sharing. Charge sharing, describing the fact that the deposited charge might be collected in more than one readout electrode, due to an angular incidence of a particle or diffusion of the generated charge carriers. It is an interesting observable for tracking detectors, as charge sharing, in combination with a charge measurement, allows for an interpolation between the centers of pixels within a cluster, which can improve the spatial resolution significantly. For a digital SiPM, this observable is expected to be dominated by optical crosstalk.

\subsubsection{Temporal Resolution}  \label{sec:analysis:obs:time}
One of the most interesting characteristics of SiPMs and SPADs is their fast response due to the small rise-time of the signals, as described in section~\ref{sec:introduction}. Similar to the spatial resolution, the temporal resolution can be determined using a reference measurement, defining a time residual
\begin{equation}
\label{eq:time_residual}
\Delta t = t_{ref} - t_{cluster}.
\end{equation}
Here, $t_{ref}$ is the time stamp provided by a reference detector, and $t_{cluster}$ is the time stamp the investigated dSiPM provides for a given cluster. This cluster timestamp is given by the earliest (smallest) pixel timestamp in the cluster, considering the two selected readout frames. The reference measurement can be taken by the TLU, i.e. the trigger time stamp, marking the first of the coinciding scintillator signals, or by the second dSiPM. In the former case, the temporal resolution of the reference measurement dominates the residual, and allows estimating only an upper limit for the temporal resolution of the dSiPM. This is an expected limitation, and the reason the setup with two dSiPMs is used.

The analysis of the corresponding temporal residuals is challenging, as they have a non-Gaussian shape. First, there is a background originating from dark counts, determining the time stamp of the quadrant TDC. The shape and width of the signal part, however, is subject to four major effects. The first two, are due to the finite effective bin width of the TDC, and bin width fluctuations, characterized by the so-called integral nonlinearity (INL). They are investigated by means of a statistical code density method, as presented in~\cite{diehl2023}. The measurements were repeated for the investigated samples. The mean bin width is found to be \SI{93.1}{\pico\second}, yielding an effective TDC resolution of \SI{26.9}{\pico\second} for the two investigated samples. The maximum INL is found to be \SI{41}{\pico\second}, but can be corrected for, by taking the individually measured bin widths into account when calculating the time stamps. The third contribution, comes from variations of the propagation delay between the pixels and the quadrant TDC. The studies with a laser setup, also discussed in~\cite{diehl2023}, reveal that this effect can be as large as $\SI{326}{\pico\second} \pm \SI{86}{\pico\second}$. It is corrected for using the measured propagation delays for each pixel, which improves the quality of the fit shown below. Otherwise, the impact of this correction is small, since the sample-to-sample variations of the propagation delay are small, and the physical alignment of the two sensors in the test beam setup is sufficiently accurate. The last contribution, comes from an inhomogeneity of the temporal response over the area of the SPAD. Evidence for that is given in~\cite{proceeding,stephan}, and further discussed in section~\ref{sec:results:temp}. Measurements with focused light in~\cite{nemallapudi2016} indicate a similar effect in analog SiPMs. This results in an exponential tail in the temporal response, which is taken into account in the analysis of the residuals.

Overall, this means that the temporal response function of a single dSiPM is characterized by three features, namely a Gaussian peak, an exponential tail due to the slow SPAD response, and a flat background. The shape of the temporal residual between the two dSiPMs is given by the convolution of two of these temporal response functions, and an empirical fit model is used to characterize the different features. The used probability density function (PDF) is
\begin{equation}
\label{eq:time_pdf}
f(\Delta t) = (1-p_{\text{noise}}) \cdot f_{\text{signal}}(\Delta t) + p_{\text{noise}} \cdot u(\Delta t),
\end{equation}
where $u(\Delta t)$ denotes the uniform distribution, which is a sufficient description of the noise contribution in the vicinity of the peak, and $p_{\text{noise}}$ determines its relative amplitude with respect to the signal PDF $f_{\text{signal}}(\Delta t)$. The signal is the sum of four additional PDFs
\begin{align}
\begin{split}
\label{eq:time_pdf_sig}
f_{\text{signal}}(\Delta t)	&= p^2_\text{fast} \cdot f_{\text{fast,fast}}(\Delta t) \\
							&+ p_\text{fast} \cdot (1-p_\text{fast}) \cdot (f_{\text{fast,slow}}(\Delta t) + f_{\text{slow,fast}}(\Delta t)) \\
							&+ (1-p_\text{fast})^2 \cdot f_{\text{slow,slow}}(\Delta t), \\
\end{split}
\end{align}
where $p_\text{fast}$ denotes the proportion of a single dSiPMs fast signal to the entire signal contribution. The PDF $f_{\text{fast,fast}}$ describes the situation where both dSiPMs exhibit a fast response, and is modeled as a Gaussian distribution with the mean $t_{\mu}$, and standard deviation of $\sigma_t \sqrt{2}$, such that $\sigma_t$ describes the standard deviation of a single dSiPMs fast response. The case where both dSiPMs are slow is described by
\begin{equation}
\label{eq:time_slow2}
f_{\text{slow,slow}}(\Delta t | \lambda) = \dfrac{\lambda}{2} \exp(-\lambda |\Delta t|).
\end{equation}
Finally there is a case where one of the dSiPM signals is fast, and the other slow. The convolution of a Gaussian and an exponential distribution, proposed in~\cite{nemallapudi2016}, is used to model this contribution. The corresponding PDF reads
\begin{equation}
\begin{split}
\label{eq:time_fastslow}
f_{\text{fast,slow}}(\Delta t | \sigma_t, \lambda,t_{\mu},t_{\text{tail}}) = \dfrac{\lambda}{2} 
	&\exp\left(\lambda(t_{\mu}+t_{\text{tail}}-\Delta t)+ \dfrac{1}{2}\lambda^2\sigma_t^2\right) \\
	&\erfc\left(\frac{t_{\mu}+t_{\text{tail}}-\Delta t + \lambda \sigma_t^2}{\sigma_t \sqrt{2}} \right).
\end{split}
\end{equation}
The PDF $f_{\text{slow,fast}}$ is defined similarly, but such that the exponential tail is on the opposite side.

An unbinned maximum likelihood estimator is used to fit the combined PDF to the temporal residuals. This approach is chosen, as it is independent of the binning of the histogram, and therefore more robust in the presence of bin-width fluctuations. Exemplary histograms and the results of the fitting procedure will be discussed in the next chapter, while more information on initial values and limits of the fit parameters are discussed in~\cite{stephan}.

It should be noted, that the data selection for the analysis of the temporal resolution is slightly different with respect to the aforementioned observables. The first difference is related to the fact, that the dead time mentioned in section~\ref{sec:analysis:obs:eff}, relates to hit detection, while there is some additional dead time while the TDCs are being reset. This leads to events with void time stamps, which are discarded for the analysis of timing properties. The second difference is that the track intercept is not required to be within the region of interest mentioned above. This increases the number of entries in the time-residual distribution, to compensate for the events discarded due to void time stamps.

		\section{Performance of the dSiPM} \label{sec:results}
The performance of a tracking detector is characterized by its capability of detecting MIPs. The key benchmarks are the detection efficiency, the spatial, and the temporal resolution.
\begin{figure}[tbp]
	\centering
	\includegraphics[width=0.5\textwidth]{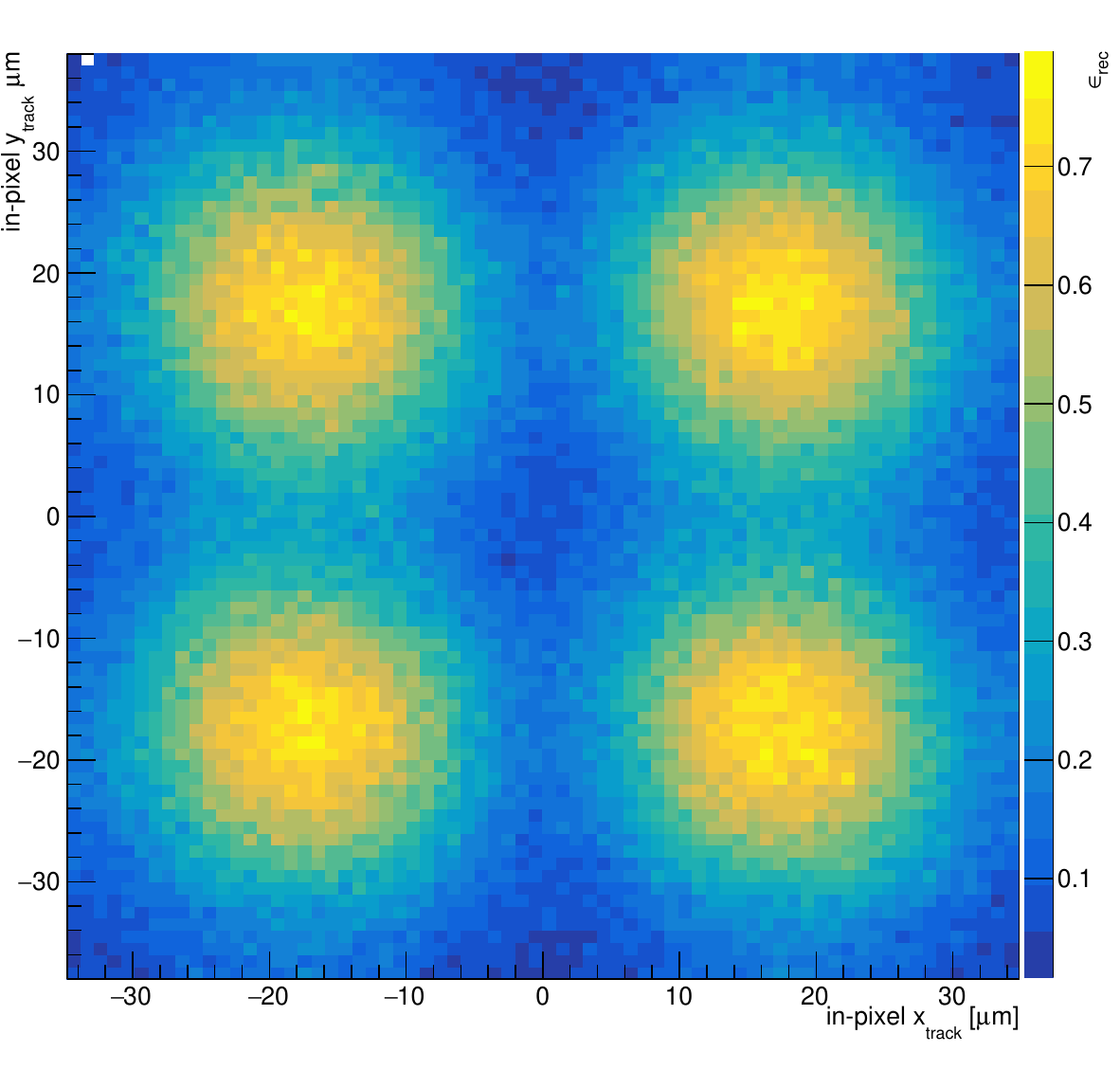}
	\caption[]{Average hit-detection efficiency as a function of the in-pixel position, measured at an overvoltage of \SI{2}{\volt}. The regions of maximum efficiency correspond to the position of the SPADs, shown in figure~\ref{fig:pixel}.}
	\label{fig:effxmym}
\end{figure}

\subsection{Hit-Detection Efficiency}
One of the major differences in the use of SPADs for MIP detection, compared to their conventional use as photodetectors, lies in the energy deposited by a MIP. While a photon in the optical wavelength regime generates a single electron-hole pair, a MIP generates more than 50 electron-hole pairs per micrometer~\cite{bichsel,wermes}. This gives raise to the assumption, that the hit-detection efficiency of the dSiPM is close to the nominal fill factor of \SI{30}{\percent}.

This is confirmed in figure~\ref{fig:effxmym}, where the reconstructed efficiency $\epsilon_{\text{rec}}$ is shown as a function of the in-pixel position, also discussed in~\cite{proceeding}. The in-pixel positions correspond to the reconstructed track intercept with the dSiPM, overlaying all selected pixels to obtain sufficient statistics. Comparing figure~\ref{fig:effxmym} with figure~\ref{fig:pixel}, shows that the efficiency reaches a maximum when the track intercepts the center of the SPAD, and a minimum in-between the SPADs. One should note that the tracking uncertainty determines the resolution of the in-pixel efficiency image. It is assumed that the efficiency in the SPAD center is actually close to \SI{100}{\percent}, and appears lower due to the smearing caused by this finite resolution.
\begin{figure}[tbp]
	\centering
	\includegraphics[width=0.5\textwidth]{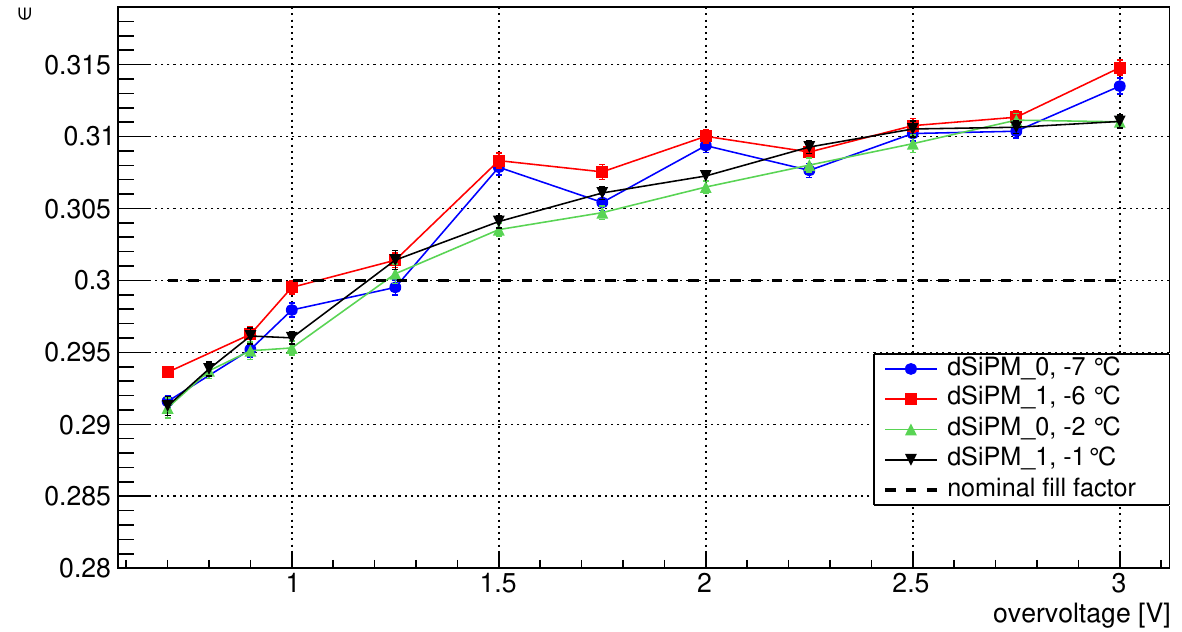}
	\caption[]{Hit-detection efficiency of two different DESY dSiPMs as a function of overvoltage, for two temperature settings, which result in slightly different chip temperatures due to different cooling contacts. The expectation from the nominal fill factor is included.}
	\label{fig:effvsov}
\end{figure}

Figure~\ref{fig:effvsov} shows the hit-detection efficiency $\epsilon$ of the two installed dSiPMs (labeled dSiPM\_0 and dSiPM\_1) as a function of the overvoltage, and for two different temperature settings. The results include the corrections introduced in equation~\ref{eq:efficiency}, and the error bars indicate statistical uncertainties only. All four measurements are found to agree within \SI{0.5}{\percent}. The small systematic uncertainties are assumed to be due to the relative movement between the dSiPMs and the beam telescope, discussed in section~\ref{sec:analysis:align}.

The hit-detection efficiency reaches a value of $\sim\SI{31}{\percent}$ at \SI{3}{\volt}. This is slightly above the nominal fill factor of \SI{30}{\percent}, and significantly higher than the expected photon-detection efficiency, due to the large amount of electron-hole pairs created by a MIP, which is assumed to cause a significant avalanche probability even when they are created in an area with a vanishing electric field. Whether the slight increase of the hit-detection efficiency with rising overvoltage is due to a growth of the active area can not be resolved due to the limited track resolution with this setup. Operating at larger overvoltages was not attempted, as further increasing signal amplitudes could damage the in-pixel electronics. For lower overvoltages, the efficiency drops sharply, as the amplitude of the signals becomes too small to pass the threshold of the inverter used as discriminator.
\begin{figure}[tbp]
	\centering
	\includegraphics[width=0.48\textwidth]{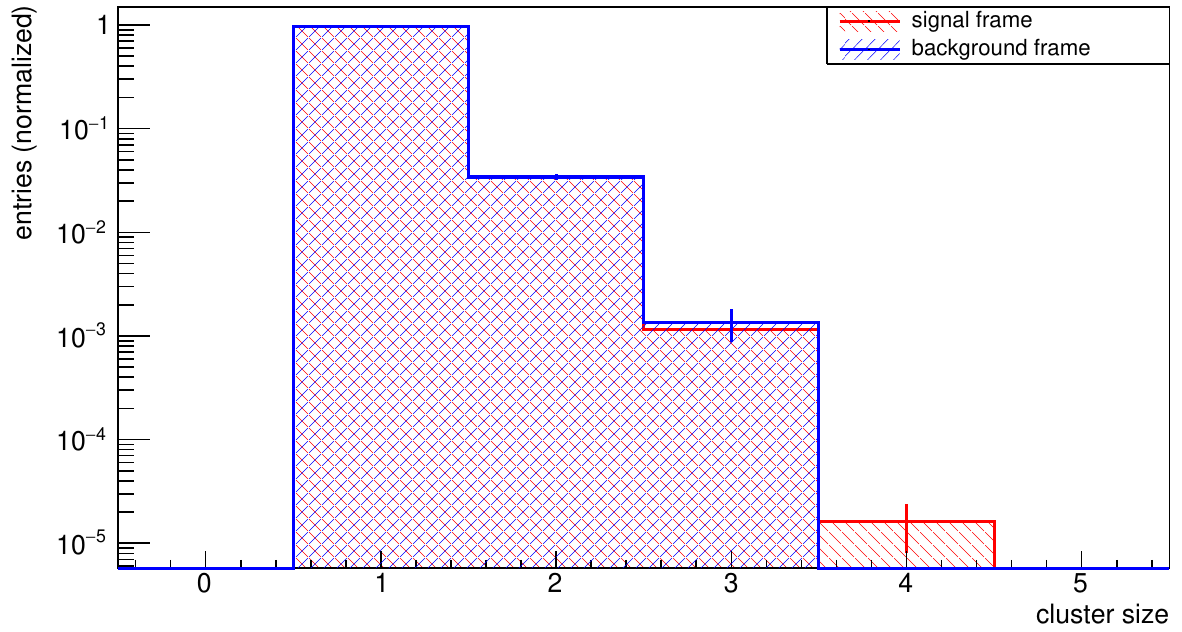}
	\caption[]{Cluster size distribution for associated cluster reconstructed within a single frame, chosen to be in time (signal frame) or out-of-time (background frame) with respect to the particle interaction. Measured at an overvoltage of \SI{2.0}{\volt} and a temperature of \SI{-2}{\degreeCelsius}.}
	\label{fig:Cluster_size}
\end{figure}

\subsection{Cluster Size and Spatial Residuals} \label{sec:results:res}
\begin{figure}[tbp]
	\centering
	\includegraphics[width=0.48\textwidth]{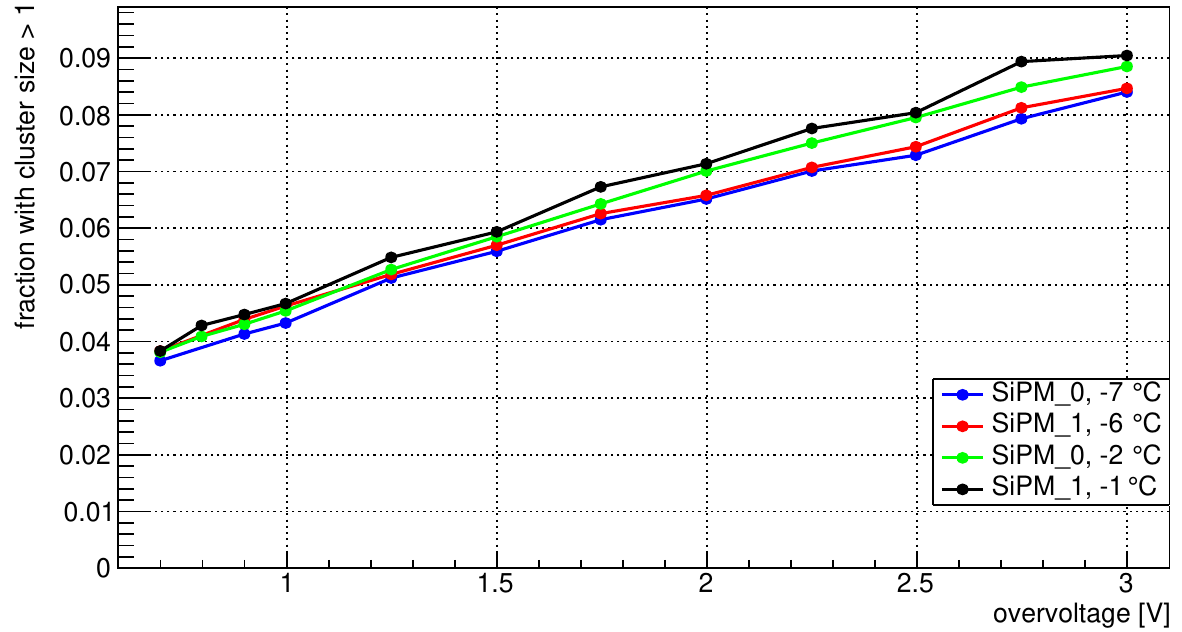}
	\caption[]{Fraction of events with cluster size greater than one, for two different DESY dSiPMs as a function of overvoltage, for two temperature settings.}
	\label{fig:Cluster_size_trend}
\end{figure}

Due to the layout of the pixels and SPADs, which are relatively wide compared to the typical thickness of the absorption region, and separated by dedicated isolation structures~\cite{diehl2023}, the cluster size for MIP interactions is expected to be close to one. This is confirmed by measurement, as shown in figure~\ref{fig:Cluster_size}, where the size of associated clusters reconstructed within a single frame is shown. This allows to compare to a second frame, which is chosen to be late with respect to the electron interaction, and hence contains predominantly falsely associated noise. The difference between the two distributions is insignificant. It is thus concluded, that the spatial distribution and amount of charge carriers generated in a MIP interaction comes with a vanishing probability of triggering an avalanche in more than one pixel. Figure~\ref{fig:Cluster_size_trend} shows the percentage of events with a DUT cluster size greater than one as a function of the overvoltage, for different temperatures and samples. It shows a marginal dependence on temperature, but increases by more than a factor of two between the lowest and highest studied overvoltage. One should note that this observable is related to crosstalk, but not corrected for the coincidental combination of hits and dark counts, or even two dark counts, hence larger than the single pixel measurements in section~\ref{sec:lab} suggest. It does not have a beneficial impact on the spatial resolution, as discussed below.

\begin{figure}[tbp]
	\centering
	\includegraphics[width=0.48\textwidth]{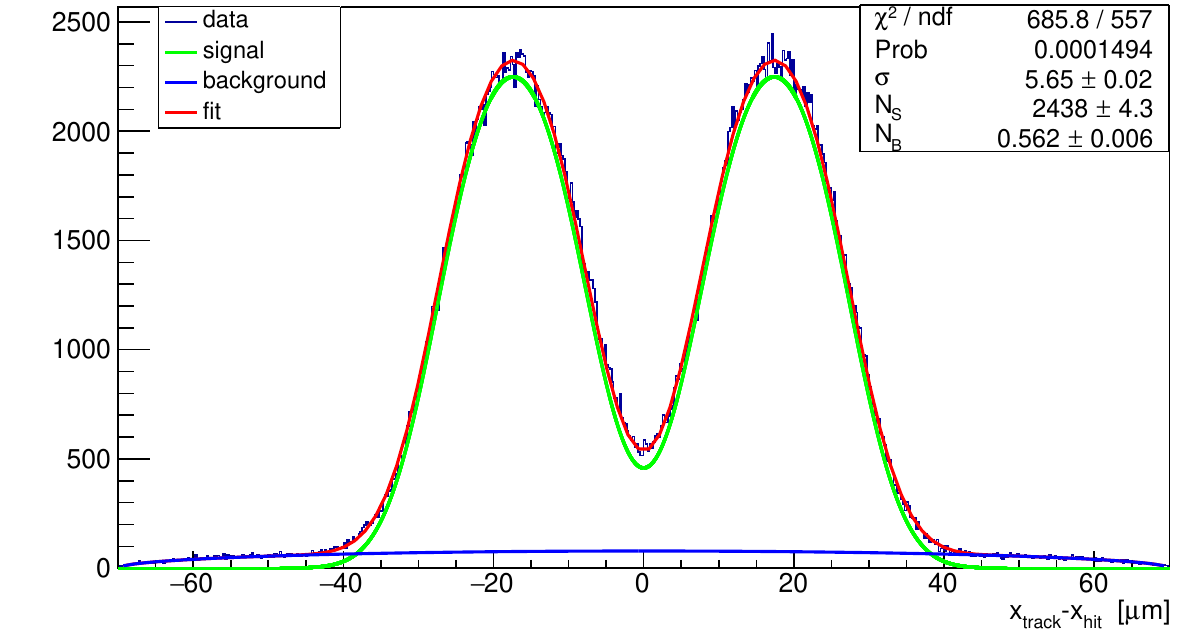}
	\caption[]{Distribution of the spatial residual $dx$, measured at an overvoltage of \SI{2}{\volt} and a temperature of \SI{-2}{\degreeCelsius}. The fitted sum of the signal $S(dx)$ and background $B(x)$ parameterization, as well as the individual contributions, are also shown.}
	\label{fig:residual}
\end{figure}

Figure~\ref{fig:residual} shows the distribution of the spatial residual $dx$, defined in equation~\ref{eq:residual}, recorded at an overvoltage of \SI{2}{\volt} and a temperature of \SI{-2}{\degreeCelsius}. The distribution assumes a characteristic shape, with a few entries around zero, due to the inefficient region between the SPADs in the pixel center, which is also apparent in figure~\ref{fig:effxmym}. The sum of the signal $S(dx)$ and background $B(x)$ parameterizations, introduced in section~\ref{sec:analysis:obs:eff}, is fitted to the data and also shown in the figure. The shape of the background distribution is validated by analyzing data from a frame which is chosen to be late with respect to the electron interaction, as done to obtain the cluster size distribution for background events. This allows to determine the standard deviation of the signal contribution, which is found to be \SI{19.2}{\micro\meter}. This result is found to be robust with respect to the fitting procedure. The main sources of uncertainty is a signal contribution to the background $B(x)$ stemming from two pixel clusters. An upper boundary for the latter effect is given by \SI{20.9}{\micro\meter}, the RMS of the full distribution in figure~\ref{fig:residual}. Taking a track resolution of the reference system of \SI{3.5}{\micro\meter} into account, yields a range of \SI{18.9}{\micro\meter} to \SI{20.6}{\micro\meter}. This is in agreement with the expectation for a sensor with binary readout, without charge sharing, and the given pitch of $\SI{69.6}{\micro\meter} / \sqrt{12} = \SI{20.1}{\micro\meter}$. Finally, the trend observed in figure~\ref{fig:Cluster_size_trend} suggests a certain dependence of the spatial resolution on the overvoltage. It is found that the spatial resolution deteriorates by about \SI{15}{\percent} from the smallest to the largest overvoltage, which is assumed to be due to the fact that the increase in cluster size is dominantly due to noise effects.

\subsection{Temporal Residuals} \label{sec:results:temp}
\begin{figure}[tbp]
	\centering
	\includegraphics[width=0.48\textwidth]{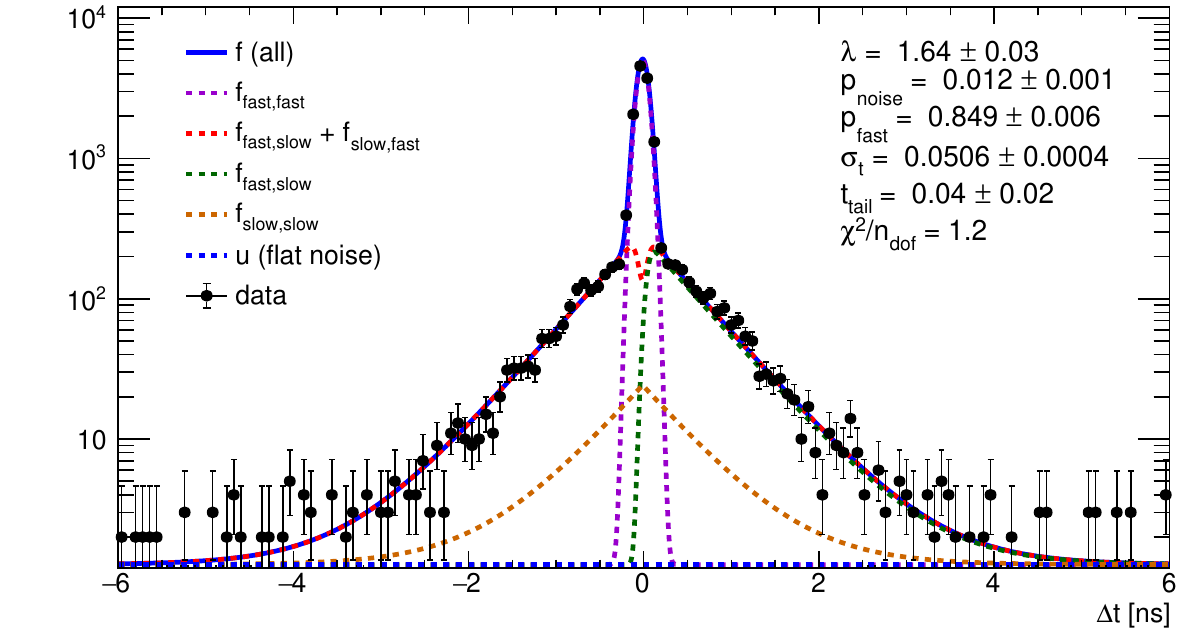}
	\caption[]{Distribution of the temporal residual between the two dSiPMs, recorded at an overvoltage of \SI{1.5}{\volt}, and temperatures of \SI{-7}{\degreeCelsius} and \SI{-6}{\degreeCelsius}. The fitted PDF and selected subcomponents, as well as key fit parameters, are shown. The unbinned likelihood fit is constrained to the displayed range, the bin width corresponds to \SI{80}{\nano\second}.}
	\label{fig:time_good}
\end{figure}

\begin{figure}[tbp]
	\centering
	\includegraphics[width=0.48\textwidth]{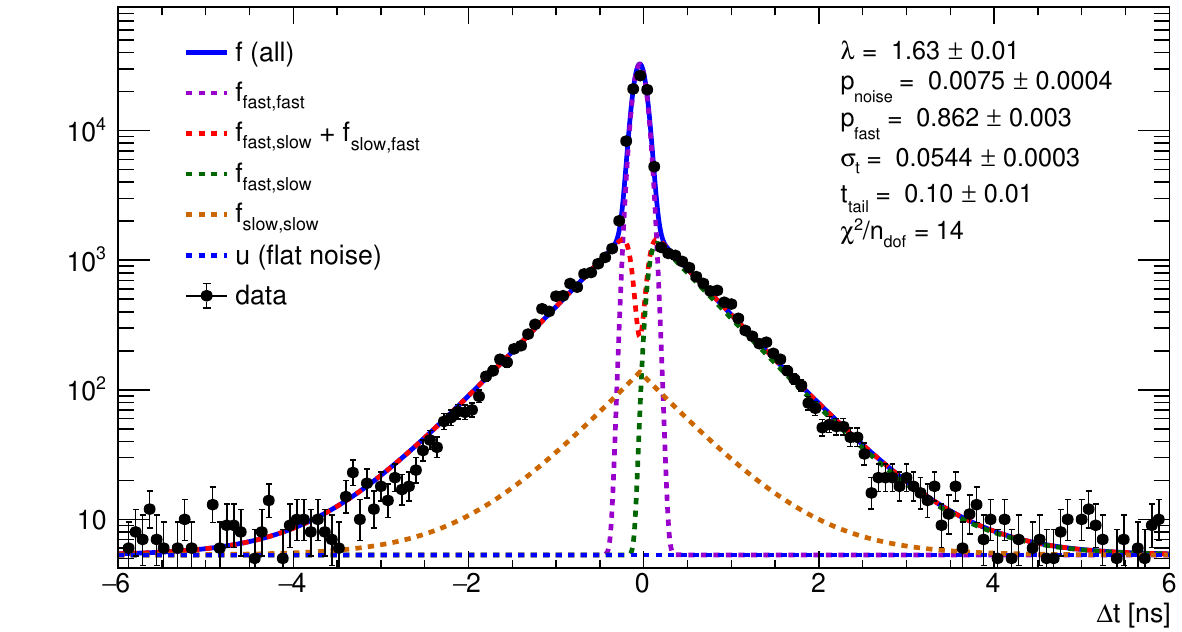}
	\caption[]{Distribution of the temporal residual between the two dSiPMs, recorded at an overvoltage of \SI{0.9}{\volt}, and temperatures of \SI{-7}{\degreeCelsius} and \SI{-6}{\degreeCelsius}. The fitted PDF and selected subcomponents, as well as key fit parameters, are shown. The unbinned likelihood fit is constrained to the displayed range, the bin width corresponds to \SI{80}{\nano\second}.}
	\label{fig:time_bad}
\end{figure}

The model and fitting procedure introduced in section~\ref{sec:analysis:obs:time} allow quantifying the temporal resolution of the dSiPM. Figure~\ref{fig:time_good} shows an exemplary distribution of the residual between the time stamps provided by the two dSiPMs, including the different contributions from the fit model. The fit results indicate, that the fraction of the fast signal amounts to about \SI{85}{\percent} of the total signal. The temporal resolution $\sigma$ of this fast part is \SI{51}{\pico\second}, while the exponential tails are characterized by time constant $\lambda = \SI{1.64}{\per\nano\second}$. The reduced chi-squared $\chi^2 / n_{\text{dof}} = 1.2$ of this fit is the best for all measurements. The fit with the worst value, $\chi^2 / n_{\text{dof}} = 14$, is shown in figure~\ref{fig:time_bad}. This might hint towards shortcomings of the empirical fit model, uncovered by the significantly smaller statistical uncertainties.

\begin{figure}[tbp]
	\centering
	\includegraphics[width=0.48\textwidth]{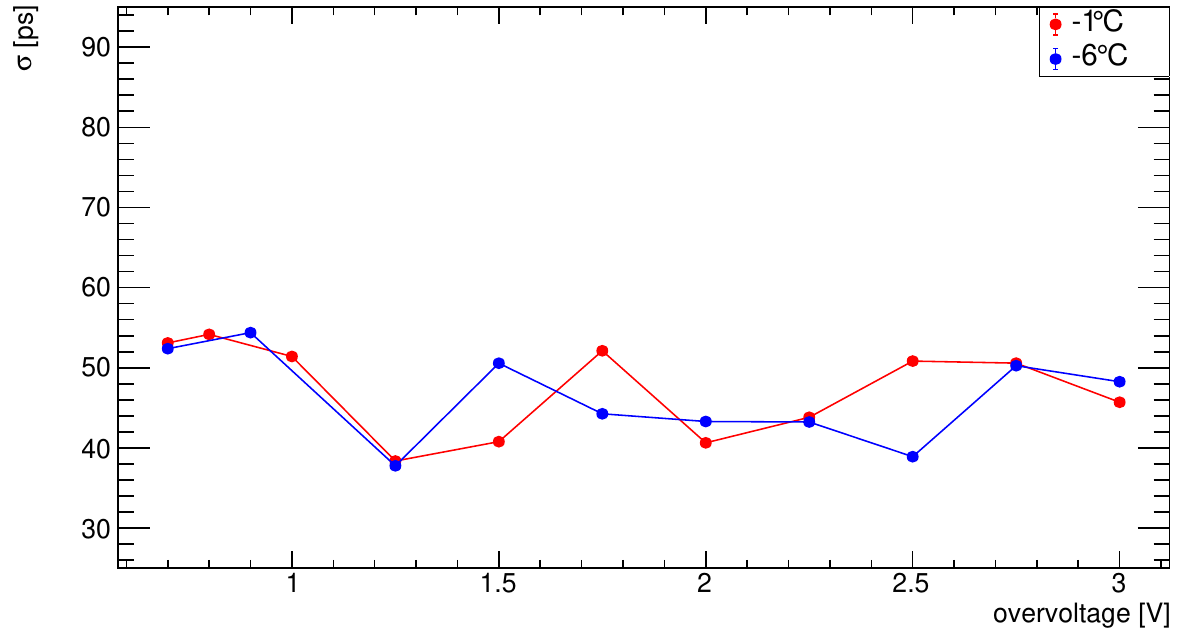}
	\caption[]{Fit parameter $\sigma$ for two different temperatures, and as a function of the overvoltage. This corresponds to the temporal resolution of the dSiPM in about \SI{85}{\percent} of the data.}
	\label{fig:timeresolution_OV}
\end{figure}

Figure~\ref{fig:timeresolution_OV} shows the width of the central peak for two different temperatures, and as a function of the overvoltage. No trend with respect to these two parameters is observed. However, the data shows fluctuations which are significantly larger than the statistical uncertainties derived from the fit. This could be due to run-to-run variations of the phase between the clock signals. As a consequence, the mean over all data points --- $\SI{47}{} \pm  \SI{6}{\pico\second}$ --- is taken as the best estimate of the dSiPMs temporal resolution. The other fit parameters are evaluated in the same fashion. The time constant $\lambda$ shows a flat behavior within \SI{1.51}{\nano\second} and \SI{1.64}{\nano\second}, while the peak fraction $p_{\text{peak}}$ is within \SI{83}{} to \SI{87}{\percent}. Only the noise fraction varies as a function of temperature and overvoltage, as could be expected from the previous section. It is found to be within \SI{0.5}{} to \SI{3.5}{\percent}.
\begin{figure}[tbp]
	\centering
	\includegraphics[width=0.48\textwidth]{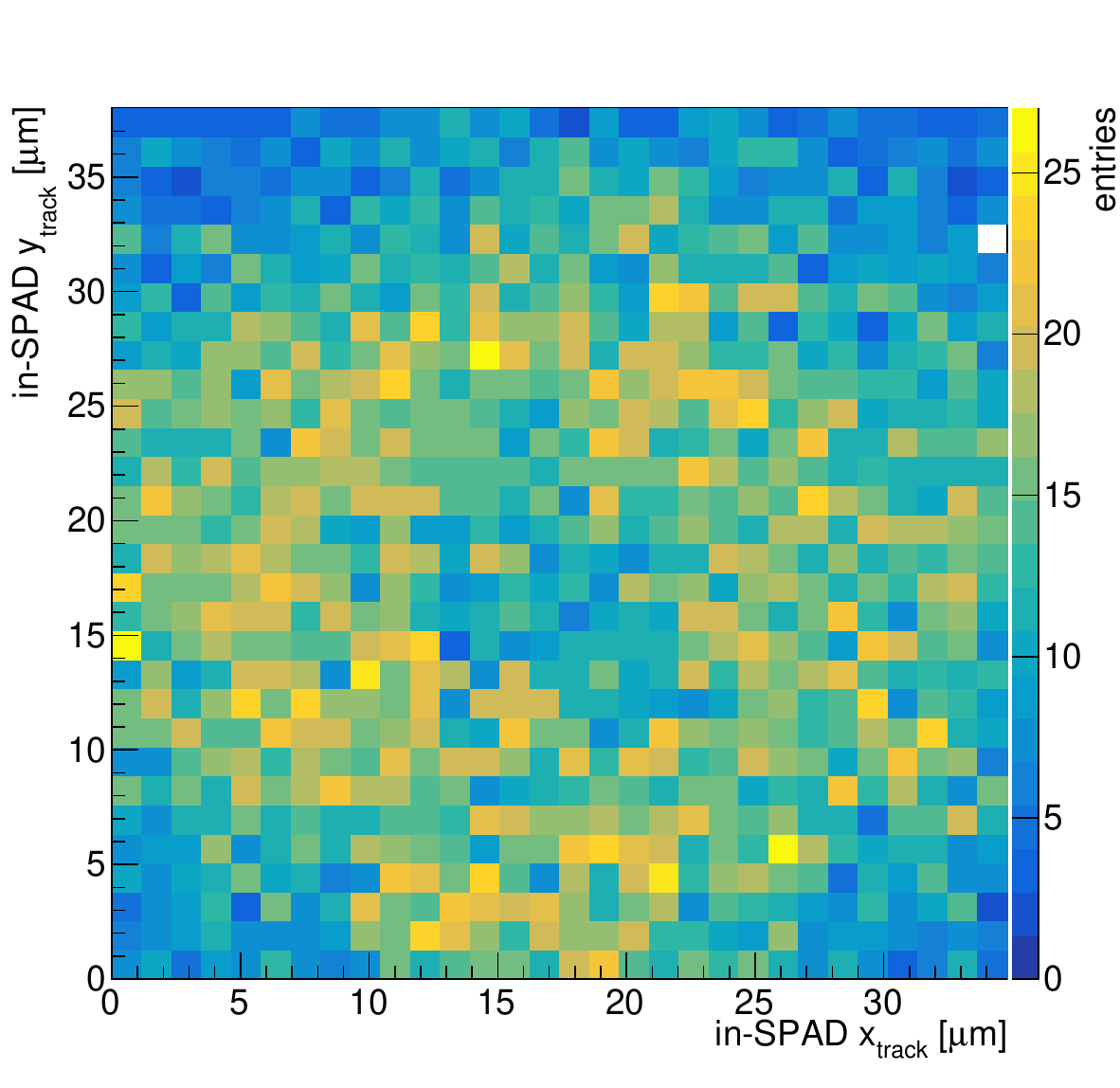}
	\caption[]{Number of events with a delay between \SI{1}{\nano\second} and \SI{5}{\nano\second} as a function of the intercepts of the reconstructed particle tracks with the dSiPM. Projected into a single SPAD to account for reduced statistics. Taken from~\cite{proceeding}.}
	\label{fig:ring}
\end{figure}

The origin of the slow component responsible for the exponential tail was studied in~\cite{proceeding,stephan}. Events with a delay between \SI{1}{\nano\second} and \SI{5}{\nano\second} with respect to the trigger time-stamp are selected, and the intercepts of the reconstructed particle tracks with the dSiPM are folded into a single SPAD (similar to figure~\ref{fig:effxmym}). The result is shown in figure~\ref{fig:ring} and indicates that most of the selected events come from a ring than corresponds to the outer region of the SPADs. The reason is assumed to be a low electric-field strength in the edge region of the SPADs, which means that charge carriers generated there require some time to reach the amplification region and to trigger an avalanche.
		\section{Conclusion and Outlook}
This work focuses on the test-beam characterization of a dSiPM prototype, designed at DESY in a standard 150nm CMOS technology, but includes also a discussion of peculiarities of the experimental methods. The results allow an assessment of the MIP detection capability of the prototype. It is shown that the MIP detection efficiency is on the order \SI{31}{\percent}, only weakly depending on the overvoltage, and dominated by the fill factor of the prototype. The spatial resolution is found to be around \SI{20}{\micro\meter}, in agreement with expectation. Studies on crosstalk reveal that the large number of charge carriers generated by a MIP comes with an insignificant increase in crosstalk probability. In addition, it is concluded, that allowing to mask individual SPADs would allow a similar reduction of the DCR compared to masking entire pixels, which would reduce the total masked area. Studies on timing performance indicate a fast response, with a temporal resolution of $\SI{47}{} \pm  \SI{6}{\pico\second}$ for about \SI{85}{\percent} of the detected events. The remaining \SI{15}{\percent} exhibit a slower response, order of nanoseconds, which correlates with particles hitting the edge of the SPADs.

Regarding the prospects of applying the investigated techology in charged particle tracking or track timing, the reported MIP detection efficiency needs to be improved. This could be achieved by increasing the fill factor, moving to larger SPADs, and narrower SPAD isolation, or even charge focusing, as presented in~\cite{canon}. A different approach is investigated in~\cite{lyso_proceeding}, where thin (few \SI{100}{\micro\meter}) scintillating crystals are coupled to the same dSiPM prototype, to improve the MIP detection efficiency through the generation of additional photons.
		\section{Acknowledgments}
Measurements leading to some of the presented results have been performed at the Test Beam Facility at DESY Hamburg (Germany), a member of the Helmholtz Association (HGF).

The authors would like to thank A. Venzmer, E. Wüstenhagen and D. Gorski for test-board and
mechanical case design, test setup, as well as chip assembly.

	\bibliography{mybibfile}

\begin{thebibliography}{10}
\expandafter\ifx\csname url\endcsname\relax
  \def\url#1{\texttt{#1}}\fi
\expandafter\ifx\csname urlprefix\endcsname\relax\def\urlprefix{URL }\fi
\expandafter\ifx\csname href\endcsname\relax
  \def\href#1#2{#2} \def\path#1{#1}\fi

\bibitem{nemallapudi2016}
M.~Nemallapudi, S.~Gundacker, P.~Lecoq, et~al., Single photon time resolution
  of state of the art {SiPMs}, J. Instrum. 11~(10) (2016) P10016.
\newblock \href {https://doi.org/10.1088/1748-0221/11/10/P10016}
  {\path{doi:10.1088/1748-0221/11/10/P10016}}.

\bibitem{acerbi2019}
F.~Acerbi, S.~Gundacker, Understanding and simulating {SiPMs}, Nucl. Instrum.
  Methods Phys. Res. A 926 (2019) 16--35, {Silicon Photomultipliers:
  Technology, Characterisation and Applications}.
\newblock \href {https://doi.org/10.1016/j.nima.2018.11.118}
  {\path{doi:10.1016/j.nima.2018.11.118}}.

\bibitem{simon2019}
F.~Simon, Silicon photomultipliers in particle and nuclear physics, Nucl.
  Instrum. Methods Phys. Res. A 926 (2019) 85--100, {Silicon Photomultipliers:
  Technology, Characterisation and Applications}.
\newblock \href {https://doi.org/10.1016/j.nima.2018.11.042}
  {\path{doi:10.1016/j.nima.2018.11.042}}.

\bibitem{bisogni2019}
M.~G. Bisogni, A.~{Del Guerra}, N.~Belcari, Medical applications of silicon
  photomultipliers, Nucl. Instrum. Methods Phys. Res. A 926 (2019) 118--128,
  {Silicon Photomultipliers: Technology, Characterisation and Applications}.
\newblock \href {https://doi.org/10.1016/j.nima.2018.10.175}
  {\path{doi:10.1016/j.nima.2018.10.175}}.

\bibitem{frach2009}
T.~Frach, G.~Prescher, C.~Degenhardt, et~al., The digital silicon
  photomultiplier --- {P}rinciple of operation and intrinsic detector
  performance (2009) 1959--1965{2009 IEEE Nuclear Science Symposium Conference
  Record (NSS/MIC)}.
\newblock \href {https://doi.org/10.1109/NSSMIC.2009.5402143}
  {\path{doi:10.1109/NSSMIC.2009.5402143}}.

\bibitem{torilla2023}
G.~Torilla, J.~Minga, P.~Brogi, et~al., {DCR} and crosstalk characterization of
  a bi-layered 24 × 72 {CMOS SPAD} array for charged particle detection,
  Nucl. Instrum. Methods Phys. Res. A 1046 (2023) 167693.
\newblock \href {https://doi.org/10.1016/j.nima.2022.167693}
  {\path{doi:10.1016/j.nima.2022.167693}}.

\bibitem{fischer2022}
P.~Fischer, R.~K. Zimmermann, B.~Maisano, {CMOS} {SPAD} sensor chip for the
  readout of scintillating fibers, Nucl. Instrum. Methods Phys. Res. A 1040
  (2022) 167033.
\newblock \href {https://doi.org/10.1016/j.nima.2022.167033}
  {\path{doi:10.1016/j.nima.2022.167033}}.

\bibitem{cartiglia2022}
N.~Cartiglia, R.~Arcidiacono, M.~Costa, et~al., {4D} tracking: present status
  and perspectives, Nucl. Instrum. Methods Phys. Res. A 1040 (2022) 167228.
\newblock \href {https://doi.org/10.1016/j.nima.2022.167228}
  {\path{doi:10.1016/j.nima.2022.167228}}.

\bibitem{diehl2023}
I.~Diehl, K.~Hansen, T.~Vanat, et~al., Monolithic {\SI{}{\mega\hertz}}-frame
  rate digital {SiPM-IC} with sub-{\SI{100}{\pico\second}} precision and
  {\SI{70}{\micro\meter}} pixel pitch, J. Instrum. 19~(01) (2024) P01020.
\newblock \href {https://doi.org/10.1088/1748-0221/19/01/P01020}
  {\path{doi:10.1088/1748-0221/19/01/P01020}}.

\bibitem{vanat2020}
T.~Vanat, {Caribou – A versatile data acquisition system}, PoS TWEPP2019
  (2020) 100.
\newblock \href {https://doi.org/10.22323/1.370.0100}
  {\path{doi:10.22323/1.370.0100}}.

\bibitem{caribou}
{Caribou developers}, \href{https://gitlab.cern.ch/Caribou}{{The Caribou DAQ
  System}}, accessed June 2024.
\newline\urlprefix\url{https://gitlab.cern.ch/Caribou}

\bibitem{amd}
\href{https://www.amd.com}{{Advanced Micro Devices, Inc.}}, accessed July 2024.
\newline\urlprefix\url{https://www.amd.com}

\bibitem{ahlburg2020}
P.~Ahlburg, S.~Arfaoui, J.-H. Arling, et~al., {EUDAQ} --- a data acquisition
  software framework for common beam telescopes, J. Instrum. 15~(01) (2020)
  P01038.
\newblock \href {https://doi.org/10.1088/1748-0221/15/01/P01038}
  {\path{doi:10.1088/1748-0221/15/01/P01038}}.

\bibitem{EUDAQ}
E.~developers, \href{https://github.com/eudaq/eudaq}{{EUDAQ code repository}},
  accessed March 2024.
\newline\urlprefix\url{https://github.com/eudaq/eudaq}

\bibitem{klanner2019}
R.~Klanner, Characterisation of {SiPMs}, Nucl. Instrum. Methods Phys. Res. A
  926 (2019) 36--56, {Silicon Photomultipliers: Technology, Characterisation
  and Applications}.
\newblock \href {https://doi.org/10.1016/j.nima.2018.11.083}
  {\path{doi:10.1016/j.nima.2018.11.083}}.

\bibitem{acs}
\href{https://www.acstestchambers.com}{{Angelantoni Test Technologies}},
  accessed February 2024.
\newline\urlprefix\url{https://www.acstestchambers.com}

\bibitem{keithley}
\href{https://www.tek.com/en/products/keithley}{{Keithley, a Tektronix
  company}}, accessed February 2024.
\newline\urlprefix\url{https://www.tek.com/en/products/keithley}

\bibitem{xtalk2013}
L.~Gallego, J.~Rosado, F.~Blanco, et~al., Modeling crosstalk in silicon
  photomultipliers, J. Instrum. 8~(5) (2013).
\newblock \href {https://doi.org/10.1088/1748-0221/8/05/P05010}
  {\path{doi:10.1088/1748-0221/8/05/P05010}}.

\bibitem{trench2009}
D.~McNally, V.~Golovin, Review of solid state photomultiplier developments by
  {CPTA} and photonique {SA}, Nucl. Instrum. Methods Phys. Res. A 610~(1)
  (2009) 150--153, {New Developments In Photodetection NDIP08}.
\newblock \href {https://doi.org/10.1016/j.nima.2009.05.140}
  {\path{doi:10.1016/j.nima.2009.05.140}}.

\bibitem{liu2016}
Z.~Liu, S.~Gundacker, M.~Pizzichemi, et~al., In-depth study of single photon
  time resolution for the {P}hilips digital silicon photomultiplier, J.
  Instrum. 11~(06) (2016) P06006.
\newblock \href {https://doi.org/10.1088/1748-0221/11/06/P06006}
  {\path{doi:10.1088/1748-0221/11/06/P06006}}.

\bibitem{dcr_lumi}
E.~Engelmann, E.~Popova, S.~Vinogradov, Spatially resolved dark count rate of
  {SiPMs}, Eur. Phys. J. C 78~(971) (2018) 8.
\newblock \href {https://doi.org/10.1140/epjc/s10052-018-6454-0}
  {\path{doi:10.1140/epjc/s10052-018-6454-0}}.

\bibitem{zeiss}
\href{www.zeiss.com}{{ZEISS Microscopy}}, accessed April 2024.
\newline\urlprefix\url{www.zeiss.com}

\bibitem{jenoptik}
\href{www.jenoptik.com}{Jenoptik}, accessed March 2024.
\newline\urlprefix\url{www.jenoptik.com}

\bibitem{desyii}
R.~Diener, J.~Dreyling-Eschweiler, H.~Ehrlichmann, et~al., The {DESY II} test
  beam facility, Nucl. Instrum. Methods Phys. Res. A 922 (2019) 265--286.
\newblock \href {https://doi.org/10.1016/j.nima.2018.11.133}
  {\path{doi:10.1016/j.nima.2018.11.133}}.

\bibitem{moliere}
G.~Moli\`{e}re, {Theorie der Streuung schneller geladener Teilchen II ---
  Mehrfach- und Vielfachstreuung}, Z. Naturforsch. Teil A 3a (1947) 78--97.
\newblock \href {https://doi.org/10.1515/zna-1948-0203}
  {\path{doi:10.1515/zna-1948-0203}}.

\bibitem{bethe}
H.~A. Bethe, Moli\`ere's theory of multiple scattering, Phys. Rev. 89 (1953)
  1256--1266.
\newblock \href {https://doi.org/10.1103/PhysRev.89.1256}
  {\path{doi:10.1103/PhysRev.89.1256}}.

\bibitem{eudet}
H.~Jansen, S.~Spannagel, J.~Behr, et~al., {Performance of the EUDET-type beam
  telescopes}, EPJ Techn. Instrum. 3~(7) (2006) 1--20.
\newblock \href {https://doi.org/10.1140/epjti/s40485-016-0033-2}
  {\path{doi:10.1140/epjti/s40485-016-0033-2}}.

\bibitem{mimosa}
C.~Hu-Guo, J.~Baudot, G.~Bertolone, et~al., First reticule size {MAPS} with
  digital output and integrated zero suppression for the {EUDET-JRA1} beam
  telescope, Nucl. Instrum. Methods Phys. Res. A 623~(1) (2010) 480--482, {1st
  International Conference on Technology and Instrumentation in Particle
  Physics}.
\newblock \href {https://doi.org/10.1016/j.nima.2010.03.043}
  {\path{doi:10.1016/j.nima.2010.03.043}}.

\bibitem{tlu}
P.~Baesso, D.~Cussans, J.~Goldstein, {The AIDA-2020 TLU: a flexible trigger
  logic unit for test beam facilities}, J. Instrum. 14~(09) (2019) P09019.
\newblock \href {https://doi.org/10.1088/1748-0221/14/09/P09019}
  {\path{doi:10.1088/1748-0221/14/09/P09019}}.

\bibitem{corry}
D.~Dannheim, K.~Dort, L.~Huth, et~al., {Corryvreckan: a modular 4D track
  reconstruction and analysis software for test beam data}, J. Instrum. 16~(03)
  (2021) P03008.
\newblock \href {https://doi.org/10.1088/1748-0221/16/03/p03008}
  {\path{doi:10.1088/1748-0221/16/03/p03008}}.

\bibitem{corry_web}
{The Corryvreckan Authors}, \href{http://cern.ch/corryvreckan}{Corryvreckan –
  {The Maelstrom for Your Test Beam Data}}, accessed January 2024.
\newline\urlprefix\url{http://cern.ch/corryvreckan}

\bibitem{gbl1}
V.~Blobel, C.~Kleinwort, F.~Meier, Fast alignment of a complex tracking
  detector using advanced track models, Comput. Phys. Commun. 182~(9) (2011)
  1760--1763, computer Physics Communications Special Edition for Conference on
  Computational Physics Trondheim, Norway, June 23-26, 2010.
\newblock \href {https://doi.org/10.1016/j.cpc.2011.03.017}
  {\path{doi:10.1016/j.cpc.2011.03.017}}.

\bibitem{gbl2}
C.~Kleinwort, General broken lines as advanced track fitting method, Nucl.
  Instrum. Methods Phys. Res. A 673 (2012) 107--110.
\newblock \href {https://doi.org/10.1016/j.nima.2012.01.024}
  {\path{doi:10.1016/j.nima.2012.01.024}}.

\bibitem{ect}
H.~Jansen, P.~Schütze, {Feasibility of track-based multiple scattering
  tomography}, Applied Physics Letters 112~(14) (2018) 144101.
\newblock \href {https://doi.org/10.1063/1.5005503}
  {\path{doi:10.1063/1.5005503}}.

\bibitem{highland}
G.~R. Lynch, O.~I. Dahl, Approximations to multiple {Coulomb} scattering, Nucl.
  Instrum. Methods Phys. Res. B 58~(1) (1991) 6--10.
\newblock \href {https://doi.org/10.1016/0168-583X(91)95671-Y}
  {\path{doi:10.1016/0168-583X(91)95671-Y}}.

\bibitem{res_calc}
S.~Spannagel, H.~Jansen, \href{https://dx.doi.org/10.5281/zenodo.48795}{{GBL
  Track Resolution Calculator v2.0}}.
\newline\urlprefix\url{https://dx.doi.org/10.5281/zenodo.48795}

\bibitem{proceeding}
I.~Diehl, F.~Feindt, K.~Hansen, et~al., {The DESY digital silicon
  photomultiplier: Device characteristics and first test-beam results}, Nucl.
  Instrum. Methods Phys. Res. A 1064 (2024) 169321.
\newblock \href {https://doi.org/10.1016/j.nima.2024.169321}
  {\path{doi:10.1016/j.nima.2024.169321}}.

\bibitem{stephan}
S.~Lachnit, \href{https://bib-pubdb1.desy.de/record/602203}{Time resolution of
  a fully-integrated digital silicon photo-multiplier}, Master's thesis,
  University of Hamburg (2024).
\newline\urlprefix\url{https://bib-pubdb1.desy.de/record/602203}

\bibitem{bichsel}
H.~Bichsel, Straggling in thin silicon detectors, Rev. Mod. Phys. 60 (1988)
  663--699.
\newblock \href {https://doi.org/10.1103/RevModPhys.60.663}
  {\path{doi:10.1103/RevModPhys.60.663}}.

\bibitem{wermes}
H.~Kolanoski, N.~Wermes, Particle Detectors: Fundamentals and Applications,
  Oxford University Press, 2020.

\bibitem{canon}
K.~Morimoto, J.~Iwata, Shinohara, et~al., {3.2 Megapixel 3D-Stacked Charge
  Focusing SPAD for Low-Light Imaging and Depth Sensing} (2021)
  20.2.1--20.2.42021 {IEEE} International Electron Devices Meeting ({IEDM}).
\newblock \href {https://doi.org/10.1109/IEDM19574.2021.9720605}
  {\path{doi:10.1109/IEDM19574.2021.9720605}}.

\bibitem{lyso_proceeding}
I.~Diehl, F.~Feindt, I.-M. Gregor, et~al., {4D-Tracking with Digital SiPMs},
  Nucl. Instrum. Methods Phys. Res. A 1069 (2024) 169985.
\newblock \href {https://doi.org/10.1016/j.nima.2024.169985}
  {\path{doi:10.1016/j.nima.2024.169985}}.

\end{thebibliography}

\end{document}